\documentclass[aps, prd, onecolumn, preprint]{revtex4}
\usepackage{graphicx}

\begin{document}

\title{Relativistic Mass Ejecta from
Phase-transition-induced Collapse of Neutron Stars
}
\author{K. S. Cheng}
\email{hrspksc@hkucc.hku.hk}
\affiliation{Department of Physics and Center for Theoretical
and Computational Physics,
The University of Hong Kong, Pok Fu Lam Road, Hong Kong}
\author{T. Harko}
\affiliation{Department of Physics and Center for Theoretical
and Computational Physics,
The University of Hong Kong, Pok Fu Lam Road, Hong Kong}
\author{Y. F. Huang}
\affiliation{Department of Astronomy, Nanjing University, Nanjing, China}
\author{L. M. Lin}
\affiliation{Department of Physics and Institute of Theoretical Physics, The Chinese University of Hong Kong, Hong Kong, China}
\author{W. M. Suen}
\affiliation{Department of Physics and Institute of Theoretical Physics, The Chinese University of Hong Kong, Hong Kong, China}
\affiliation{McDonnell Center for the Space Sciences, Department of Physics, Washington University, St. Louis, USA}
 \author{X. L. Tian}
\affiliation{Department of Physics,
The University of Hong Kong, Pok Fu Lam Road, Hong Kong}

\begin{abstract}

We study the dynamical evolution of a phase-transition-induced collapse neutron
star to a hybrid star, which consists of a mixture of hadronic matter and strange quark matter.
The collapse
is triggered by a sudden change of equation of state, which result in a large
amplitude stellar oscillation. The evolution of the system
is simulated by using a 3D Newtonian hydrodynamic code with a high resolution
shock capture scheme. We find that both the temperature and the density at the
neutrinosphere are oscillating with acoustic frequency. However, they are nearly
180$^{\circ}$ out of phase. Consequently, extremely intense,
pulsating neutrino/antineutrino fluxes will be
emitted periodically. Since the energy and density of neutrinos at the peaks of the pulsating fluxes are much
higher than the non-oscillating case, the electron/positron pair creation rate can be enhanced
dramatically. Some mass layers on the stellar surface can be ejected by absorbing energy of neutrinos and pairs. These
mass ejecta
can be further accelerated to relativistic speeds by absorbing
electron/positron pairs, created by the neutrino and antineutrino
annihilation outside the stellar surface. The possible connection between this process and
the cosmological Gamma-ray Bursts is discussed.

\end{abstract}


\date{\today}

\maketitle

\section{Introduction}

Gamma-ray bursts (GRBs) are cosmic gamma ray emissions with typical
fluxes of the order of $10^{-5}$ to $5\times 10^{-4}$ergs cm$^{-2}$
with the rise time as low as $10^{-4}$ s and the duration of bursts
from $10^{-2}$ to $ 10^{3}$ s. The distribution of the bursts is
isotropic and they are believed to have a cosmological origin, with
the observations showing that GRBs originate at extra-galactic
distances. The large inferred distances imply isotropic energy
losses as large as $3\times 10^{53} $ ergs for GRB 971214 and
$3.4\times 10^{54}$ ergs for GRB 990123. For detailed reviews of
GRBs properties see \cite{Me06} and \cite{Zh07}, respectively. The
widely accepted interpretation of the phenomenology of GRBs is that
the observable effects are due to the dissipation of the kinetic
energy of a relativistically expanding fireball, whose primal cause
is not yet known \cite{Me06}.

One fundamental question related to GRBs is how many intrinsically
different categories they have. Each type may correspond to an
intrinsically different type of progenitor, as well as to a
different type of central engine. From the GRB sample collected by
BATSE, a clear bimodal distribution of bursts was identified
\cite{Kou93,Zh07}. Two criteria have been used to classify the
bursts. The primary criterion is duration, and a separation line of
2 s has been adopted to separate the double-hump duration
distribution of the BATSE bursts. The second criterion is the
hardness-usually denoted as the hardness ratio within the two energy
bands of the detector. On average, short GRBs are harder, while long
GRBs are softer. Hence the observations show the existence of two
classes of GRBs, long-soft, and short-hard, respectively.
The short GRBs are hard because of
the harder low-energy spectral index of the GRB spectral function \cite{Gh04}.
More interestingly, short GRB spectra are broadly similar to those
of long GRBs, if only the first 2 seconds of data of the long GRBs
are taken into account. Afterglows observation shed some light into
the physical nature of these two types of GRBs. The host galaxies of
long GRBs are exclusively star forming galaxies, predominantly
irregular dwarf galaxies \cite{Fru06}. The observations suggest
that most, if not all of the long GRBs are produced during the core
collapse of massive stars, called collapsars, as has been also
suggested theoretically \cite{Wo93, Pa98, Mac99}. Both the
observations and the theoretical models also support the idea that
long GRBs are associated with supernova explosions \citep{WoBl06}.

The observations of SWIFT and of HETE led to a very different
picture for the short GRBs \cite{Me06, Zh07}. The basic result of
all these observations is that short GRBs are intrinsically
different from the long GRBs. Most of the short GRBs are found at
the outskirts of elliptical galaxies, where the star forming rate is
very low. Even if they can be associated with star forming regions,
they are rather far away from the active zones \cite{Bl06}. In
several cases a robust host galaxy has not been identified, but the
host galaxy is of early type. Deep supernova searches have been
performed, but with negative results \cite{Zh07}. This suggests
that short GRBs may be associated with mergers of compact objects,
like neutron star-neutron star or neutron star-black hole mergers,
white-dwarf-neutron star or black hole mergers etc. \cite{Pa86,
RuJa99, DeAt06, Le06, Ki07}.

The possibility that the conversion of neutron stars to strange quark
stars may be the energy source for the cosmological gamma ray bursts
was suggested in \cite{ChDa96}. Neutron stars in low-mass X-ray
binaries can accrete sufficient mass to undergo a phase transition
to become strange stars. At the moment of its birth the strange star
is very hot, with an interior temperature of around $\sim 10^{11}$
K. By approximating the strange matter by a free Fermi gas, the
thermal energy of the star is $E_{th}\approx 5\times
10^{51}\left(\rho /\rho _0\right)^{2/3}R_6^3T_{11}^2$ ergs, where
$\rho $ is the average mass density, $\rho _0=2.8\times 10^{14}$
g/cm$^3$ is the nuclear density, $R_6$ is the radius of the star in
units of $10^6$ cm, and $T_{11}$ is the temperature in units of
$10^{11}$ K. For $\rho =8\rho _0$, $R_6=1$, $T_{11}=1.5$ the thermal
energy of the newly formed quark star is $E_{th}\approx 5\times
10^{52} {\rm ergs}$. In the original model of \cite{ChDa96} it was
assumed that the star would cool by emission of neutrinos and
antineutrinos, and that the neutrino pair annihilation process $\nu
\bar{\nu}\rightarrow e^-e^{+}$ operates near the strange star
surface. The energy deposited due to this process is $E_{1}\approx
2\times 10^{48}\left( T_{0}/10^{11}K\right) ^{4}{\rm ergs}\approx
10^{49}$ ergs, where $T_{0}$ is the initial temperature. The time
scale for the deposition is around 1 s. On the other hand, the
processes $n+\nu _{e}\rightarrow p+e^{-}$ and
$p+\bar{\nu}_{e}\rightarrow n+e^{+}$ play an important role in the
energy deposition. The integrated neutrino optical depth due to all
these processes is $\tau \approx 4.5\times 10^{-2}\rho
_{11}^{4/3}T_{11}^{2}$, where $\rho _{11}$ is the mass density in
units of $10^{11}$ g/cm$^{3}$. The deposition energy can be
estimated to be $E_{2}\approx E_{th}\left(1-e^{-\tau }\right)
\approx 2\times 10^{52}$ ergs. Here the value of the neutron drip
density, $\rho _{11}=4.3$ has been used, and it has been assumed
that all the thermal energy of the star is lost in neutrinos. The
process $\gamma \gamma \rightarrow e^-e^{+}$ inevitably leads to the
creation of a fireball,  and this fireball will expand outward. The
expanding shell interacts with the surrounding interstellar medium,
and its kinetic energy is radiated through non-thermal processes in
shocks. However, this model did not consider any internal dynamics,
e.g. heat transport, viscous damping, shock dissipation etc, which
can affect the neutrino emission dramatically. In fact the numerical
simulations indicate that the temperature on the neutrinosphere is
rapidly changing with time, steady neutrino emission is not
possible.

The presence of oscillations of the resulting quark star produced by
the phase transition induced collapse of a neutron star is one of
the most intriguing features of the simulations performed with our
Newtonian numerical code introduced in \cite{Lin06}. Recently,
the collapse process with a
conformally flat approximation to general relativity was also simulated in \cite{Ab08}. The works of
\cite{Lin06,Ab08} focus on the gravitational wave signals emitted
by the collapse process.

It is the purpose of the present paper to
consider another important implication of this result, namely, the
effect of the oscillations of the newly formed quark star on the
neutrino emission. The oscillations can enhance the neutrino
emission rate in a pulsating manner, and the neutrinos are emitted
in a much shorter time scale. Therefore the neutron-quark phase
transition in compact objects may be the energy source of GRBs.

This paper is organized as follows. The phase transition process
from neutron stars to hybrid stars, which consists of a mixture of strange quark
matter and hadronic matter, is summarized in Section II. We
describe our numerical code, which is used to simulate the dynamical
evolution of star after phase transition in Section III. In Section IV,
we calculate the neutrino and antineutrino emission from the
neutrinosphere and the electron/positron pair creation rate due to neutrino
and antineutrino annihilation process. In Section V, we calculate
mass ejection from stellar surface by absorbing neutrinos/pairs and their
subsequent acceleration by the pairs outside the star. In
Section VI we apply our model to GRBs. Finally a brief summary and
discussion is presented in Section VII.

\section{Description of the phase transition}

The quark structure of the nucleons, suggested by quantum chromodynamics,
indicates the possibility of a hadron-quark phase transition at high densities
and/or temperatures, as suggested by \cite{It70, Bo71, Wi84}. Theories of the
strong interaction, like, for example, the quark bag models, suppose that breaking of physical vacuum
takes place inside hadrons. If the hypothesis of the quark matter is true,
then some of neutron stars could actually be strange stars, built entirely
of strange matter \cite{Al86, Ha86, Ch98, Ch98a, ChDa}.

The central density of compact stellar objects may reach values of
up to ten times nuclear-matter saturation density, and therefore a
phase transition to deconfined quark matter, or pion and kaon
condensates, should take place at least in the central region of the
neutron stars. In the case of the transition to quark matter, in
addition to a phase of unpaired normal quark matter, present at low
densities, several superconducting phases, such as the {\bf two-flavor
color superconductor} (2SC) phase or the gapless Color-flavor-lock
(CFL) phase can also occur at the large baryon densities reached at
the central regions of a compact star. The transition from the
hadronic to the quark phase could proceed in two steps. In the first
step, a transition from hadronic matter to normal quark matter or to
a 2SC phase takes place, due to the increase of the baryonic density
at the center of the star. This increase in density may be due to
mass accretion from the fall-back material or rapidly spin-down of
the star. The newly formed hybrid or quark star, containing some 2SC
quark matter, can become meta-stable, and decay into a star
containing a CFL phase \cite{Dr07}. {\bf Pure hadronic compact stars above a threshold value of their mass are metastable. The metastability of hadronic stars originates from the finite size effects in the formation process of the first strange quark matter drop in the hadronic environment \cite{Bom2003}.}

A phase transition between the hadronic and quark phase occurs when
the pressures and the chemical potentials in the two phases are
equal, $P_{h}=P_{q},\mu _{h}=\mu _{q}$, where $P_{h}$, $\mu _{h}$
and $P_{q}$, $\mu _{q}$ are the pressures and the chemical
potentials in the hadron and quark phase, respectively. {\bf In the present Section, unless otherwise explicitly specified, we use the natural system of units with $\hbar=c =1$}. If the
transition pressure is less than that existing at the center of the
compact object, the transition can occur. For finite values of the
surface tension, complicated structures can develop in the mixed
phase, such as drops, rods, and slabs \citep{Gl00}. The formation of
a mixed phase is the result of two competing processes: the size of
the barrier that the system has to overcome in order to form a
structure, and the size of the perturbation of the system. One
method to see if the phase transition can proceed or not is to
compare the temperature reached by the system immediately after the
conversion with the height of the barrier. If the temperature is not
much lower than the height of the barrier, the structure formation
proceeds thermally, and it is very rapid \cite{OlMa93, OlMa94}. If
the temperature is low, new structures can form only via quantum
nucleation, which is a very slow process \cite{Dr07}.

{\bf The thermal nucleation rate of quark drops can be estimated in
the framework of the nucleation theory as $R_{nucl}\approx a\exp
\left( -W_{c}/T\right) $, where the prefactor $a$ (the product of the dynamical prefactor and of the statistical prefactor) is the product of a density and a growth factor,
$W_{c}=W\left( R_{c}\right) $ represents the work needed to form the
smallest drop capable of growing, and $T$ is the temperature. $W_{c}$ corresponds to the
maximum of the free energy of the drop in the new phase. The dynamical prefactor for the nucleation rate of bubbles or droplets in first-order phase transitions for the case where both viscous damping and thermal dissipation are significant has been obtained in \cite{VeVi94}. This formalism was applied for the study of the nucleation of quark-gluon plasma from hadronic matter in \cite{KaViVe95}. By taking into account the explicit forms of the temperature dependent quark matter equation of state, of the dissipative factors and of the statistical prefactor, one obtains a complicated dependence of the prefactor on the temperature. However, the prefactor in this model is based on poorly known physical parameters (like, for example, the shear viscosity of the hadronic phase). In order to obtain some qualitative estimates of the nucleation rate we use the thermodynamical approach developed in \cite{OlMa93, OlMa94}. The form of the prefactor can be obtained from general thermodynamical considerations as $a\approx T^4$, and it is very little dependent on some particular choices (for example in the low temperature limit the prefactor in the expression for $R_{nucl}$ can be replaced by the baryon chemical potential). This form for the prefactor does not include any kinematics, e.g. the microscopic processes required to transform a gas of hadrons into a gas of quarks. In the expression of the nucleation rate the dominant term is the exponential.}
The free
energy is given
by $W=-4\pi R^{3}\Delta P/3+4\pi \sigma R^{2}+8\pi \gamma R+N_{q}\Delta \mu $%
, where $\Delta P=P_{q}-P_{h}$ is the pressure difference, $\sigma
=\sigma _{q}+\sigma _{h}$ is the surface tension, $\gamma =\gamma
_{q}-\gamma _{h}$ is the curvature energy density and $\Delta \mu
=\mu _{q}-\mu _{h}$ is the difference in the chemical potential.
$N_{q}$ is the total baryon number in the quark drop
\cite{OlMa93, OlMa94, Ha04, Dr07}.

The free energy has a maximum at the critical radius $%
R_{c}=\sigma \left( 1+\sqrt{1+b}\right) /C$, where $C=\Delta
P-n_{q}\Delta
\mu $ and $b=2\gamma C/\sigma ^{2}$. The corresponding free energy is
\begin{equation}
W_{c}=8\pi \sigma ^{3}\left[ 1+\left( 1+b\right) ^{3/2}+3b/2\right] /3C^{2}.
\end{equation}
The
number $N$ of drops of the new phase formed inside the old phase in
a volume $V$ in a time $t$ is given by $N=R_{nucl}Vt$. Let $\lambda
$ be the spacing between two drops in the mixed phase. The number
of drops in a volume $V$ is given by $V/\lambda ^{3}$. The number
of drops produced while the front moves over a distance $ \lambda
$ must be of the order of the drops that are present in the mixed
phase, $R_{nucl}Vt=R_{nucl}S\left( \lambda /v\right) \geq V/\lambda
^{3}=S/\lambda ^{2}$, where $v$ is the velocity of the front and $S$
its surface area. Therefore in order for thermal nucleation take
place the condition $W_{c}/T\leq \ln \left( T^{4}\lambda
^{4}/v\right) $ must be satisfied. The process of absorption of a
hadron into a pure quark matter phase can also be described
phenomenologically as the fusion of a small drop of quarks growing
into a much larger drop.

{\bf The nature of the conversion process from the neutron to the quark matter is not yet fully understood, and there is no clear theoretical evidence if it is a deflagration or a detonation. Indeed, for realistic equations of state of quark and neutron matter and when matter is assumed to be in $\beta $ equilibrium, detonation is difficult to achieve \cite{HoBe88, Cho94}. However, it would be still possible to obtain a detonation if the matter immediately after the front is not yet in $\beta $-equilibrium \cite{Dr07}. It is also important to note that neutrino trapping delays $\beta $ stability. According to the analysis of \cite{Dr07}, when the drop starts expanding, the process of conversion can be extremely fast, within the layer in which deconfinement is energetically favorable, even in the absence of the weak processes. In this case, the conversion front moves at the velocity of the deflagration front $v_{df}$, which approaches the velocity of sound. As the conversion layer moves outward, the front enters the region of mixed phase, where $v_{df}$ decreases until it vanishes at the low density boundary of the mixed phase.

If the conversion is indeed extremely fast, taking place at speeds close to the speed of sound, then the typical time scale for the
transition can be estimated as $\tau _{tr}=R/c_{s}$, where $R$ is the radius of the
neutron star and $c_{s}$ is the speed of the sound.} A simple
phenomenological model for the evolution of the quark phase can be
obtained by assuming the relation $dr/dt=\left(r-R_{c}\right)/\tau
_{tr}$ \cite{Ha04}, which gives for the transition time scale
$T_{tr}$ from a microscopic quark drop to a quark matter
distribution of a macroscopic size the expression
\begin{equation}\label{eq2}
T_{tr}\approx 10^{-4}N_q^{-1/3}R_6\ln \frac{R}{ R_{q}}~{\rm s},
\end{equation}
where $R_{6}$ is the neutron star radius in units of $10^6$ cm, $R_q
\sim 300R_c$ is the initial size of the quark
drop \cite{Ha04}, and $N_q$, which could be as large as $10^{48}$ \cite{IS98},
is the number of quark drops inside the core of the neutron star.
In this case $R$ in Eq.~(\ref{eq2}) is replaced by $R/N_q^{1/3}$.
Therefore the phase transition may take place in a time scale much shorter than
submillisecond.

\section{Simulation of the phase-transition induced collapse}

In this paper we focus on studying the phase-transition induced
collapse of  neutron star to a hybrid quark star, which consists of a mixture of strange
quark matter and hadronic matter, and want to
demonstrate that this process can produce relativistic ejecta, which
could be a mechanism for GRBs. In order to avoid other complications
we would like to simulate a non-rotating, phase-induced collapse of
a compact object, with a mixed phase of quark matter and nuclear
matter. In this model gravitational radiation will not be emitted.
Hence we can focus on studying the neutrino and pair emissions and
subsequently gamma-ray emission via interaction between the
interstellar medium from this system.
In our simulations, we will not simulate the phase transition process. Instead,
we assume that a fast phase transition has happened (e.g., via a detonation
mode) so that the initial neutron star has converted to a quark star in a
timescale shorter than the dynamical timescale of the system.
We assume that the normal matter inside the initial
neutron star has suddenly changed to quark matter at t=0. This is achieved
by changing the EOS at t=0 after the initial hydrostatic equilibrium neutron
star has been constructed. We then simulate the resulting dynamics of the system
triggered by the collapse.

\subsection{Description of the numerical code}

Our numerical code is based on the three-dimensional numerical
simulation in Newtonian hydrodynamics and gravity. The quark matter
of the mixed phase is described by the MIT bag model and the normal
nuclear matter is described by an ideal fluid EOS. This code has
been used to study the gravitational wave emission from the
phase-induced collapse of the neutron stars \cite{Lin06}. Here we
briefly summarize the main equations and the numerical scheme
involved. A detailed discussion can be found in \cite{Lin06}.

The system of equations describing the non-viscous Newtonian fluid
flow is given by
\begin{equation}
        \frac{\partial \rho}{\partial t} + \nabla \cdot
        \left( \rho {\bf v} \right) = 0 ,
\label{eq:rhoeq}
\end{equation}
\begin{equation}
        \frac{\partial}{\partial t} \left( \rho v_{i} \right) +
        \nabla \cdot \left( \rho v_{i} {\bf v} \right) +
        \frac{\partial P}{\partial x_{i}} = - \rho
        \frac{\partial \Phi}{\partial x_{i}} ,
\label{eq:momeq}
\end{equation}
\begin{equation}
        \frac{\partial \tau}{\partial t}
        + \nabla \cdot \left(
        \left( \tau + P \right)
        {\bf v} \right) = -\rho {\bf v} \cdot \nabla \Phi  ,
\label{eq:taueq}
\end{equation}
where $\rho$ is the mass density of the fluid, ${\bf v}$ is the
velocity with Cartesian components $v_i$ ($i=1,2,3$), $P$ is the
fluid pressure, $\Phi$ is the Newtonian potential and $\tau$ is the
total energy density, $\tau = \rho \epsilon + \rho {\bf v}^{2}/2$,
and $\epsilon$ is the internal energy per unit mass of the fluid,
respectively. The Newtonian potential $\Phi$ is obtained by solving
the Poisson equation, ${\nabla}^{2} \Phi = 4 \pi G \rho $. The
system is completed by specifying an equation of state $P=P(\rho,
\epsilon)$.

The above hydrodynamics Eqs.~(\ref{eq:rhoeq})-(\ref{eq:taueq}) can
be rewritten in the so-called flux-conservative form, which can be
solved numerically using quite standard high resolution shock
capturing (HRSC) schemes. A HRSC scheme, using either exact or
approximate Riemann solvers, with the characteristic fields
(eigenvalues) of the system, obtains the solution of a local Riemann
problem at every cell interface of a finite-differencing grid. Such
schemes have the ability to resolve discontinuities in the solution
(e.g., shock waves) by construction. Moreover, they have high
accuracy in regions where the fluid flow is smooth. In general,
integrating the hydrodynamics equations by a HRSC scheme involves
the choices of an appropriate numerical flux formula and a
reconstruction method of the state variables $(\rho,\epsilon,{\bf
v})$ for solving the Riemann problems at the cell interfaces. In our
code, we use the Roe's approximate Riemann solver \cite{roe81} for
the numerical fluxes and the third-order piecewise parabolic method
\cite{col84} for the reconstruction. For the temporal
discretization, we use a basic two-step method to achieve
second-order accuracy in time.

In the simulations, we introduced a very low density atmosphere
outside the star. The ''artificial'' atmosphere is not physical, but
it is important  for the stability of the  hydrodynamical code. This
is due to the problem that the hydrodynamical codes cannot in
general  handle vacuum  regions where the density is zero. In order
to avoid a significant influence of the atmosphere on the dynamics
of the physical system, it is necessary to choose the density of the
atmosphere $\rho_{\rm atm}$ to be much smaller than the density
scale of interest. For the results reported in Section 3.4, we set
$\rho_{\rm atm}$ to be $3\times 10^9 \;{\rm g/cm}^3$. The effects of
different atmospherical values have been compared in \cite{T08}.

\subsection{Equation of state}

The equation of state (EOS) for neutron stars is highly uncertain.
We could try all possible existing realistic EOS in our study.
However, the main purpose of this paper is to demonstrate that
during the phase-transition induced collapse of a neutron star
extremely intense, pulsating and very high energy neutrinos can be
emitted. The effect is governed mainly by the amount of pressure
reduction after the phase transition as compared to the initial
neutron star model. For simplicity we will use a polytropic EOS for
the initial equilibrium neutron star:
\begin{equation}
P=k_0\rho^{\Gamma_0} , \label{eq:poly_EOS}
\end{equation}
where $k_0$ and $\Gamma_0$ are constants. On the initial time slice,
we also need to specify the specific internal energy $\epsilon$. For
the polytropic EOS, the thermodynamically consistent $\epsilon$ is
given by
\begin{equation}
\epsilon = {k_0\over \Gamma_0-1} \rho^{\Gamma_0-1} .
\end{equation}
Note that the pressure in Eq.~(\ref{eq:poly_EOS}) can also be
written as
\begin{equation}
P=(\Gamma_0-1)\rho\epsilon .
\end{equation}

{\bf In order to describe the physical properties of the neutron stars after the phase transition, we consider that the star can be divided in three regions. At the center of the star, where $\rho >\rho _q$, we have a pure quark core (Region I), which is described by the MIT bag model equation of state, so that the pressure $P$ is given by
\begin{equation}
P=P_{\rm q} = {1\over 3} \left( \rho + \rho\epsilon - 4B\right),\rho>\rho _q,
\label{eq:quark_eos}
\end{equation}
where $B$ is the bag constant, and $\rho _q$ is a critical density for which all the hadrons are deconfined into quarks. It should be noticed that $P_q$ is not in the usual form
of $P = (\rho_{\rm tot} - 4 B)/3$, where $\rho_{\rm tot}$ is
the (rest frame) total energy density.
It is because in our Newtonian simulations, we use the rest mass
density $\rho$ and specific internal energy $\epsilon$ as
fundamental variables in the hydrodynamics equations. We assume that the quark core is absolutely stable.
The quark core is surrounded by a mixed phase of quark and nuclear matter (Region II) that can exist if the
density of the region is higher than a certain critical value $\rho_{tr}$ (quark
seeds can spontaneously produce everywhere when $\rho \geq \rho
_{tr}$).  Explicitly, the pressure in the mixed phase is given by
\begin{equation}
P =  \alpha P_{\rm q} + (1-\alpha) P_{\rm n}, \rho_q>\rho > \rho_{tr} 
\label{eq:mixed_EOS}
\end{equation}
where 
\begin{equation}
P_{\rm n}=(\Gamma_n-1)\rho \epsilon , \label{eq:eos_idealgas}
\end{equation}
 and
\begin{equation}
\alpha = \left\{ \begin{array}{cc}
          { (\rho -\rho_{tr}) / (\rho_{q} - \rho_{tr}) },
         & \ \mbox{for} \ \  \rho_{tr} < \rho < \rho_{\rm q}, \\
       \\
            1,
         & \mbox{for} \ \ \rho_{q} < \rho ,  \end{array}  \right.
\label{eq:alpha}
\end{equation}
is defined to be the scale factor of the mixed phase \citep{Lin06}.
$\Gamma_n$ is not necessarily equal to $\Gamma_0$.
Finally, we have a normal nuclear matter region (Region III), extending from $\rho <
\rho_{tr}$ to the surface of the star, so that in this region}
\begin{equation}
P=P_{\rm n},{\rm for} \ \ \rho  \leq \rho_{tr}.
\end{equation}
 A more detailed discussion
about such hybrid quark stars can be found in \cite{Lin06}.

The total
energy density $\rho_{\rm tot}$, which includes the rest mass
contribution, is decomposed as $\rho_{\rm tot} = \rho +
\rho\epsilon$. We choose $\Gamma_n < \Gamma_0$ in our simulations to
take into account the possibility that the nuclear matter may not be
stable during the phase transition process, and hence some quark
seeds could appear inside the nuclear matter, or the convection,
which can occur during the phase transition process, can mix some
quark matter with the nuclear matter. In the presence of the quark
seeds in the nuclear matter, the effective adiabatic index will be
reduced. The possible values of $B^{1/4}$ range from 145 MeV to 190
MeV \cite{DeGr75,Sa82,Ste95,Ha03}. For
$\rho
>\rho_q$, the quarks will be deconfined from nucleons. The value of
$\rho_q$ is model dependent; it could range from 4 to 8 $\rho_{nuc}$
\citep{ChDa98a,Ha03, Bo04}, where
$\rho_{nuc}=2.8\times 10^{14}~{\rm g~cm}^{-3}$ is the nuclear
density.

There are two issues regarding our EOS model to be addressed:
(1) It should be noted that, for simplicity, we do not include the change
in the internal energy from the phase transition when setting the initial
data for the collapse. The binding energy released in the phase transition
effectively leads to a slightly harder EOS due to the thermal pressure. This
could be modeled by using a larger value of $\Gamma_n$ (comparing to the one
we used in this work). However, as long as the EOS after the phase transition
is softer than that of the initial neutron star, the star will still
collapse and stellar pulsations will be triggered.
(2) Furthermore, the parameter $\rho_{tr}$ in our EOS model should be
considered as the density below which the matter is dominated by hadronic
matter. It is noted that strange quark matter (if it is more stable) cannot
convert back to hadronic matter by decreasing the density. However,
when a fluid element originally in the mixed phase moves to the lower
density region ($\rho < \rho_{tr}$), the fluid element will mix with a large
amount of hadronic matter. Hence, we approximate that the pressure in the
outer region of the star is dominated by the hadronic part, which is modeled
by an ideal gas.

In the simulations, we choose $\Gamma_0 = 2$, $\Gamma_n = 1.85$,
$B^{1/4} = 160$ MeV and $\rho_q = 9 \rho_{\rm nuc}$. The transition
density $\rho_{tr}$ is defined to be at the point where $P_q$
vanishes initially.

\begin{figure}
\includegraphics[scale=1, keepaspectratio = false ]{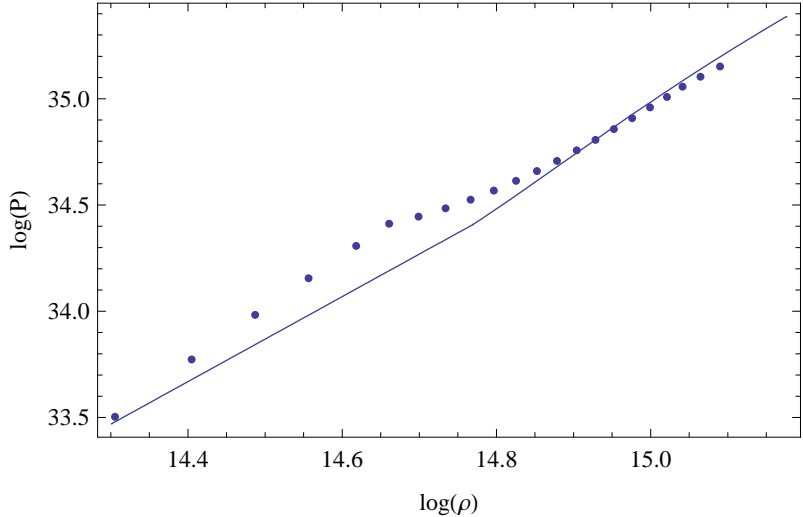}
\caption{Comparison of the equation of state of the mixed quark-hadron phase proposed in \cite{Gl00} for $K=240$ MeV and effective mass $m*/m=0.78$ (dotted curve), and Eq.~(\ref{eq:mixed_EOS}), the equation of state if the mixed phase used in the present simulations (solid curve).}
\label{pres}
\end{figure}

In Eq.~(\ref{eq:mixed_EOS}), we have used a very simple linear combination of strange quark matter EOS and hadronic matter to represent the EOS of mixed phase,
which consists of a mixture of strange quark drops and hadronic matter. In fact,
the properties of a mixed quark-hadron phase and its implications for hybrid star structure were considered in \cite{Gl00}.  In Fig.~\ref{pres} we compare the EOS of the mixed phase obtained in \cite{Gl00} for a compression modulus $K=240$ MeV and an effective mass at saturation density of $m*/m=0.78$ with Eq.~(\ref{eq:mixed_EOS}) used in the present simulation. We can see that these EOSs are quite close to each other. Although we use Eq.~(\ref{eq:mixed_EOS}) in our simulation solely based on simplicity, we believe that the simulation results will not be changed qualitatively if we replace Eq.~(\ref{eq:mixed_EOS}) by the EOS given in \cite{Gl00}.

{\bf In the mixed phase we have assumed that the effective bag constant $B$ is a linearly dependent function of the density. The equation of state used by us reproduces quite well the EOS proposed in \cite{Gl00} to describe the mixed quark-hadron phase. It is important to note that the main purpose of our paper is to focus on the study of the dynamical response of the star after a sudden phase transition, no matter what the fine details of the phase transition are. A full and exact description of the phase transition would require a very precise knowledge of the physical parameters describing both quark and hadronic matter. However, taking into account the uncertainties in our present knowledge of the properties of matter at high densities,  our investigations could certainly give at least a qualitative picture of the astrophysical implications of hadron-quark phase transitions in compact stars.}

\subsection{The Neutrinosphere}

For a new born compact star, the internal temperature is so high
that neutrinos will be trapped inside the star for at least a few
seconds (cf.~\cite{Shapiro1983} for a general review). However,
neutrinos very near the surface of the star can still escape from
the star because the optical depth near the stellar surface is low.
Quantitatively we can define a surface called the neutrinosphere
with a radius $R_{\nu}$ as follows, e.g.
\cite{Shapiro1983,Jaroszynski1996},
\begin{equation}
\label{eq:Janka2001eq16} \tau_{eff}=\int_{R_\nu}^\infty dr \,
\kappa_{eff}(r)=1,
\end{equation}
where the effective optical depth, $\tau_{eff}$ is defined as the
inverse mean free path and the effective opacity, $\kappa_{eff}$ is
given by
 \begin{equation}\label{eq:opacityFinal}
\langle\kappa_{eff}\rangle(r)=1.202\times10^{-7}\rho_{10}(r)\left(\frac{T_\nu}{4\;\textrm{MeV}}\right)
^{2}\quad\frac{1}{\textrm{cm}}.
\end{equation}
It is clear that this surface is a function of the temperature and
of the density.

\subsection{Simulation Results}

The total time span of the simulations is $\sim$5 ms, and the time
step is $3.7\times 10^{-4}$ ms. For all the simulations we report in
this paper, the grid spacing is set to be $dx=0.28$ km and the outer
boundary of the computational domain is at 27.5 km, which is about
two times the stellar radius of our models. With the grid resolution
we used for the simulations, we see that numerical damping becomes
important after about 3 ms. We shall thus only present the numerical
results up to 3 ms. During the simulations, a low-density atmosphere
is added outside the neutron star. The density and temperature of
the atmosphere are $3 \times 10^{9} \textrm{g} / \textrm{cm}^3$ and
$0.003$ MeV respectively.

\begin{figure}
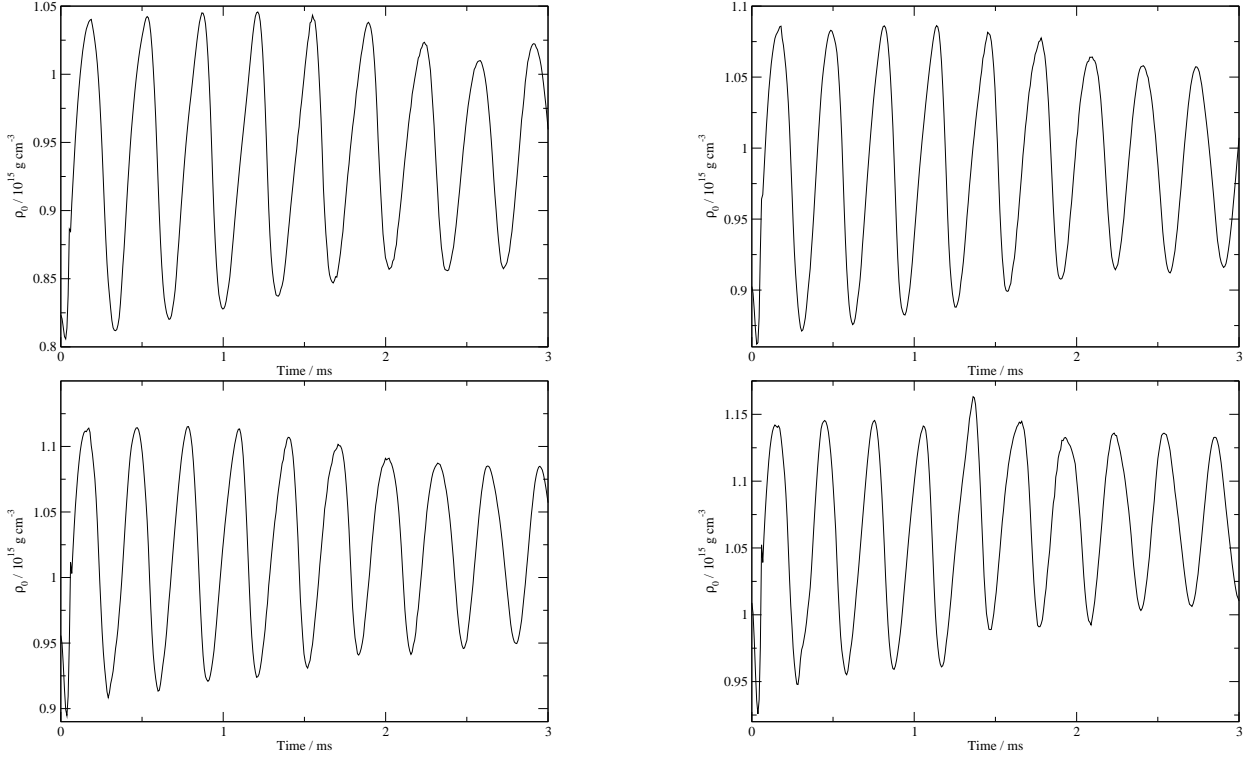

\includegraphics*[scale=0.3, keepaspectratio = false ]{fig1a.eps}
\hfill\includegraphics*[scale=0.3,keepaspectratio = false]{fig1b.eps}
\hfill\includegraphics*[scale=0.3,keepaspectratio = false]{fig1c.eps}
\hfill\includegraphics*[scale=0.3,keepaspectratio = false]{fig1d.eps}
\caption{Variation of the central density with respect to the time for
$M=1.55M_{\odot}$, (left upper figure), $M=1.7M_{\odot}$, (right upper
figure), $M=1.8M_{\odot}$, (left lower figure), $M=1.9M_{\odot}$, (right
lower figure).}
\label{fig:rhoc_time_models}
\end{figure}

Fig.~\ref{fig:rhoc_time_models} shows the oscillations of the central
density for stars with
$M=1.55, 1.7, 1.8$ and $1.9 M_{\odot}$ respectively.

We can see that the oscillation period gradually decreases from $\sim 0.32$ ms to
$\sim 0.29$ ms, when the mass of the star increases. It is because the
radial oscillation is basically acoustic oscillation (cf. Lin et al.
2006) therefore frequency essentially increases with the average
density. In fact the frequency is roughly scaled according to $\rho_0^{1/2}$.
Hence, the oscillation period is smaller for more massive
stars because they have higher density. We can also see that the
oscillation amplitudes decrease slightly after 2 ms.
We note that the damping is due partly to finite-differencing errors
and physical effects, such as mass-shedding near the surface.
One might worry that the large amplitude oscillations seen in the collapse
simulations were numerical artifacts due to numerical errors or instability.

\begin{figure}
\includegraphics[scale=0.7, keepaspectratio = false ]{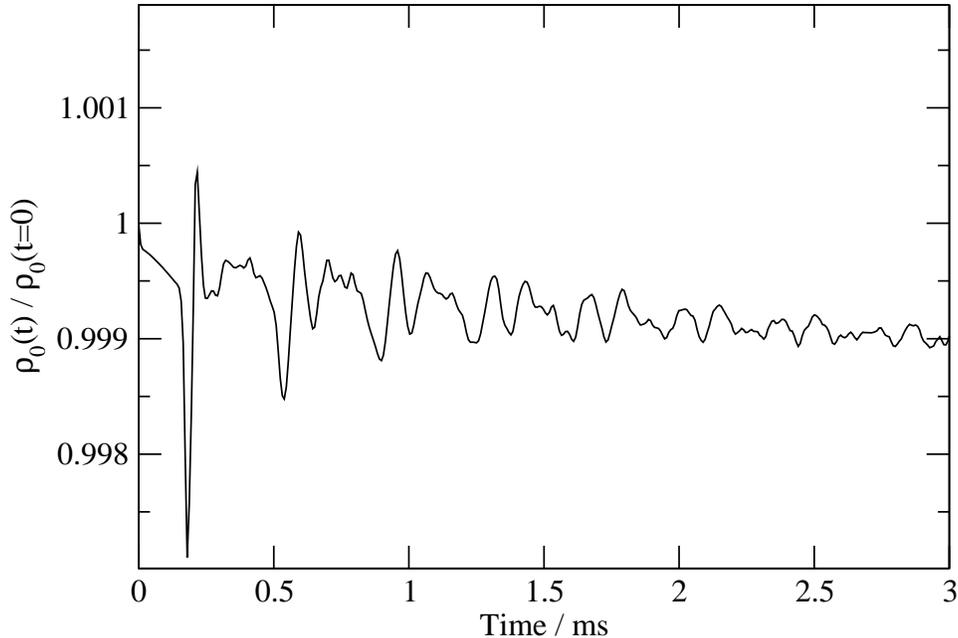}
\caption{Evolution of the central density (normalized by its initial value)
for an unperturbed equilibrium neutron star with $M=1.8M_\odot$.}
\label{fig:M1.8_unperturb_rhoc}
\end{figure}

In order to demonstrate that the oscillations were in fact triggered by the
phase transition, we show in Fig.~\ref{fig:M1.8_unperturb_rhoc} the evolution
of the central density (normalized by its initial value) for an unperturbed
equilibrium neutron star model with $M=1.8 M_\odot$.
We see that the amplitude of the oscillations of the star triggered by
finite-differencing errors is much smaller than that seen in the collapse
models. Furthermore, there is no obvious periodic features in Fig.~\ref{fig:M1.8_unperturb_rhoc}.

\begin{figure}[ht]
\includegraphics[scale=0.4, keepaspectratio = false ]{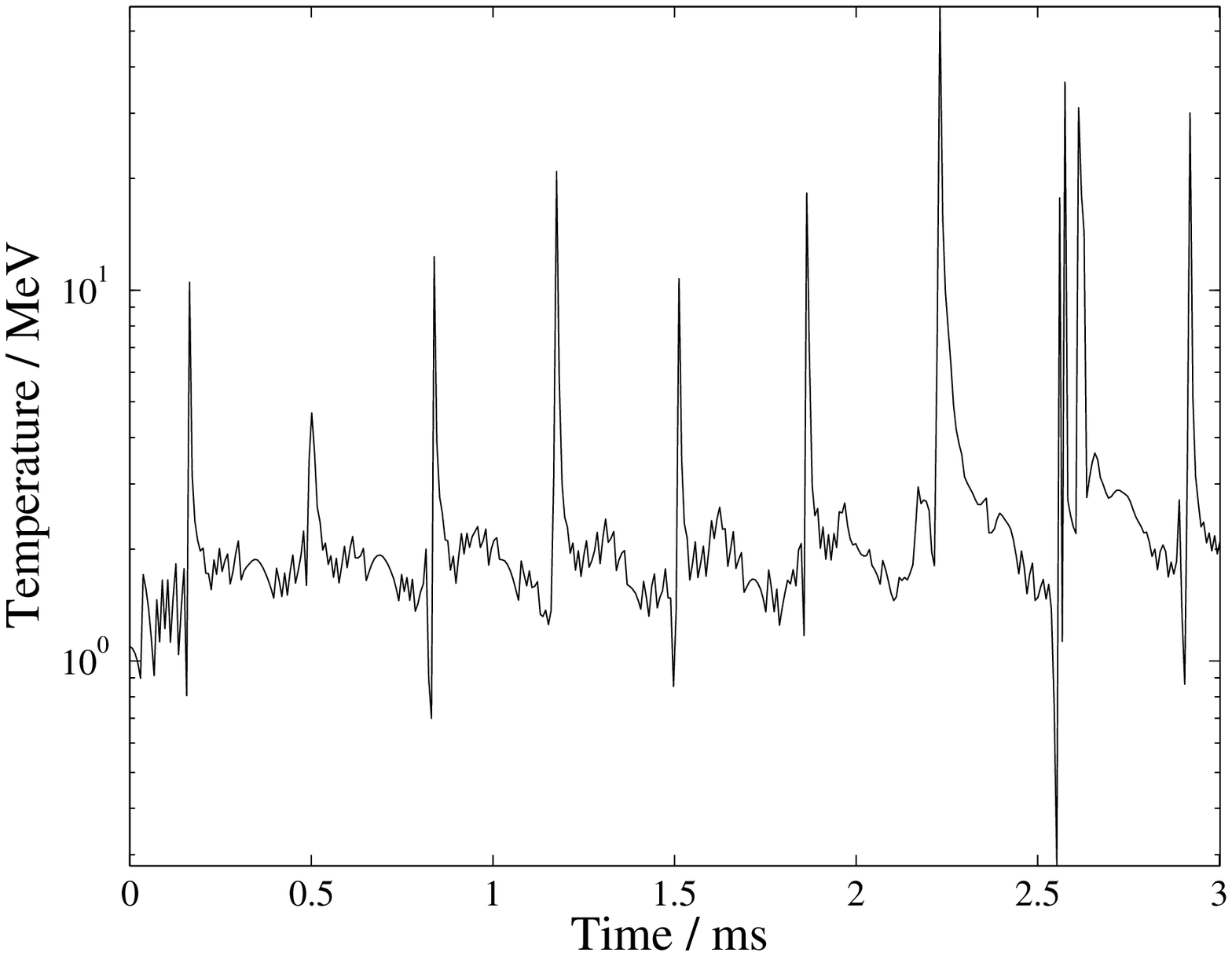}
\hfill\includegraphics[scale=0.4,keepaspectratio = false]{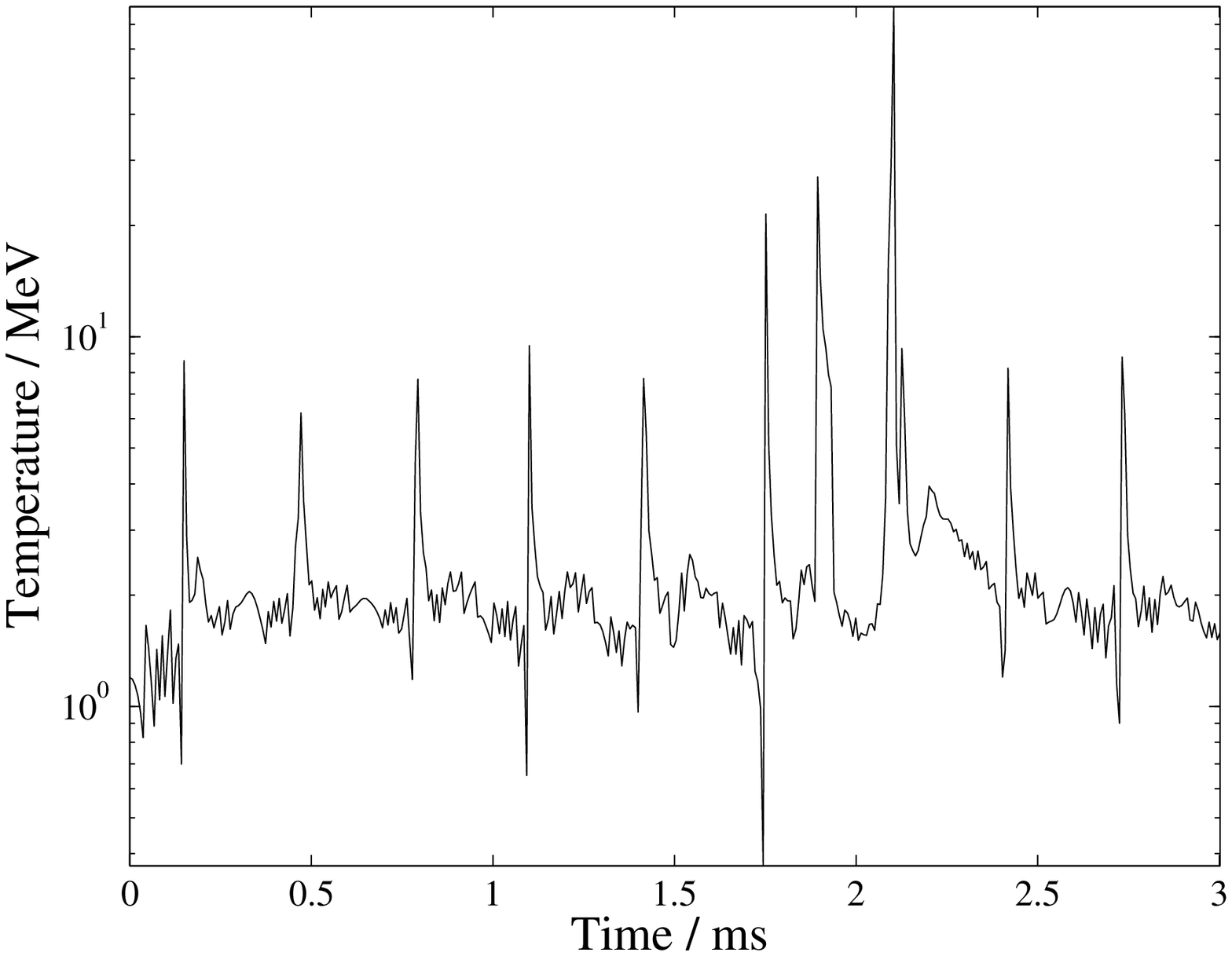}
\hfill\includegraphics[scale=0.4,keepaspectratio = false]{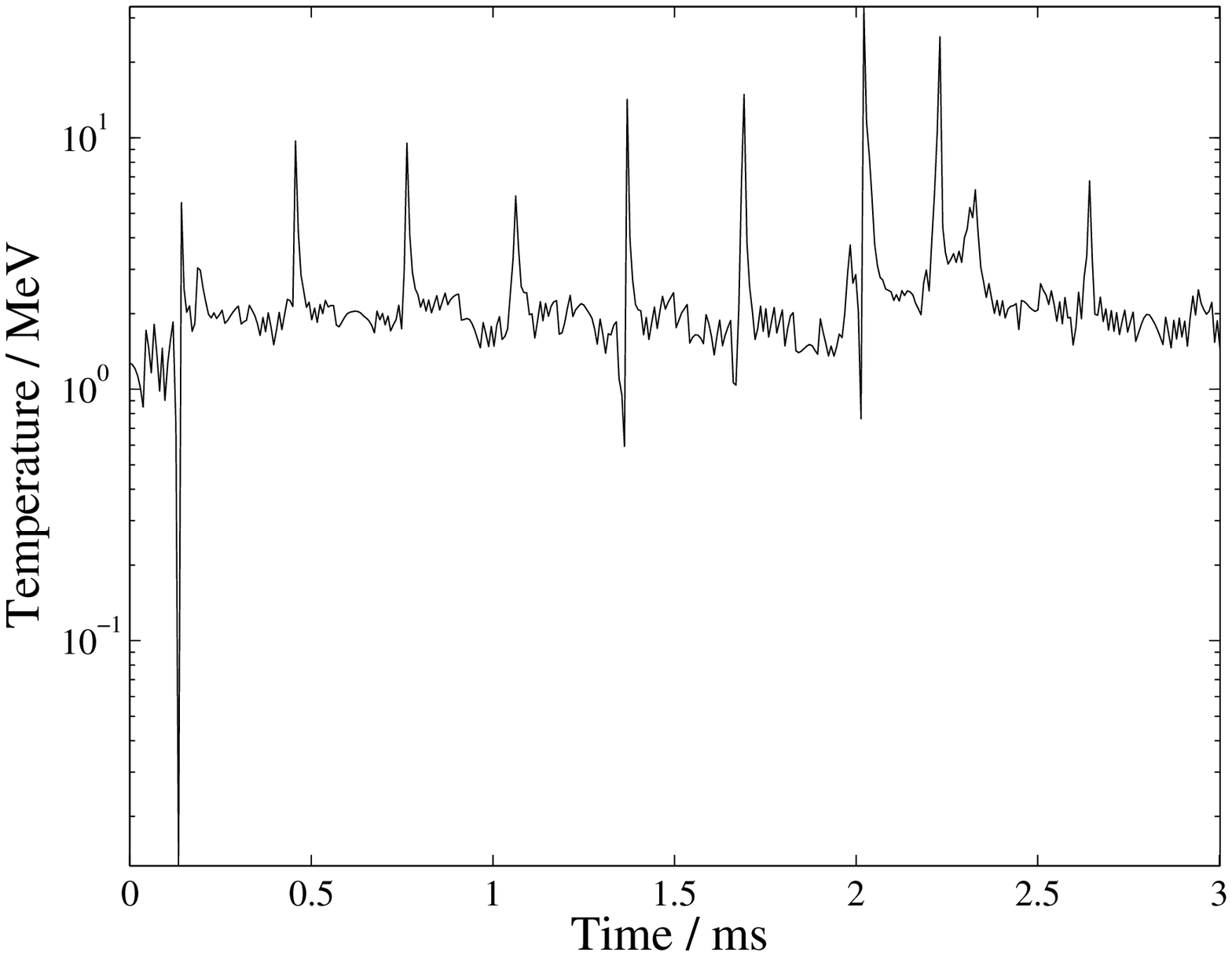}
\hfill\includegraphics[scale=0.4,keepaspectratio = false]{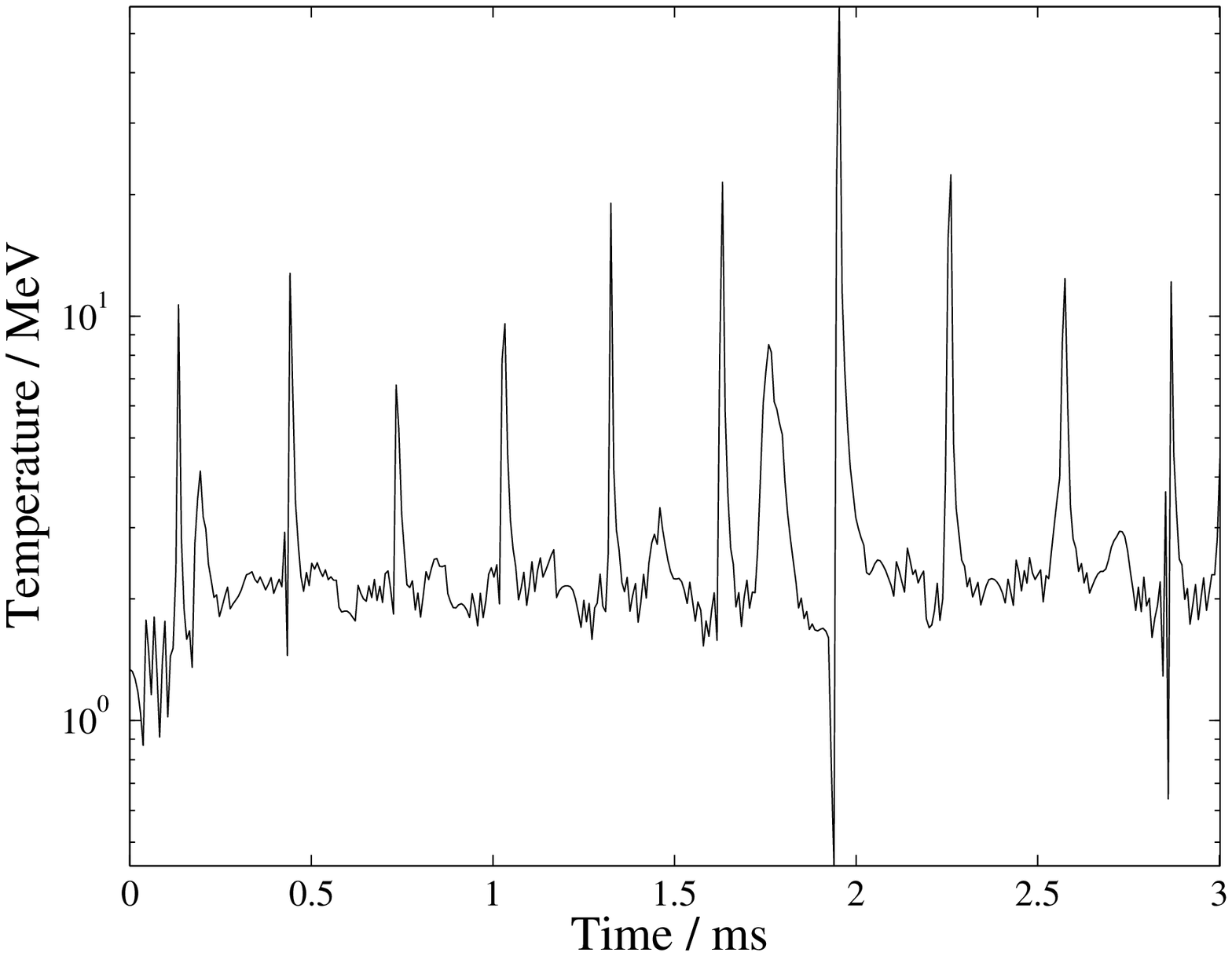}
\caption{Temperature at the neutrinosphere versus time for $M=1.55M_{\odot}$, (left upper figure), $M=1.7M_{\odot}$, (right upper figure), $M=1.8M_{\odot}$, (left lower figure), $M=1.9M_{\odot}$, (right lower figure).}
\label{fig:Tnu_time_models}
\end{figure}

\begin{figure}[ht]
\includegraphics[scale=0.4, keepaspectratio = false ]{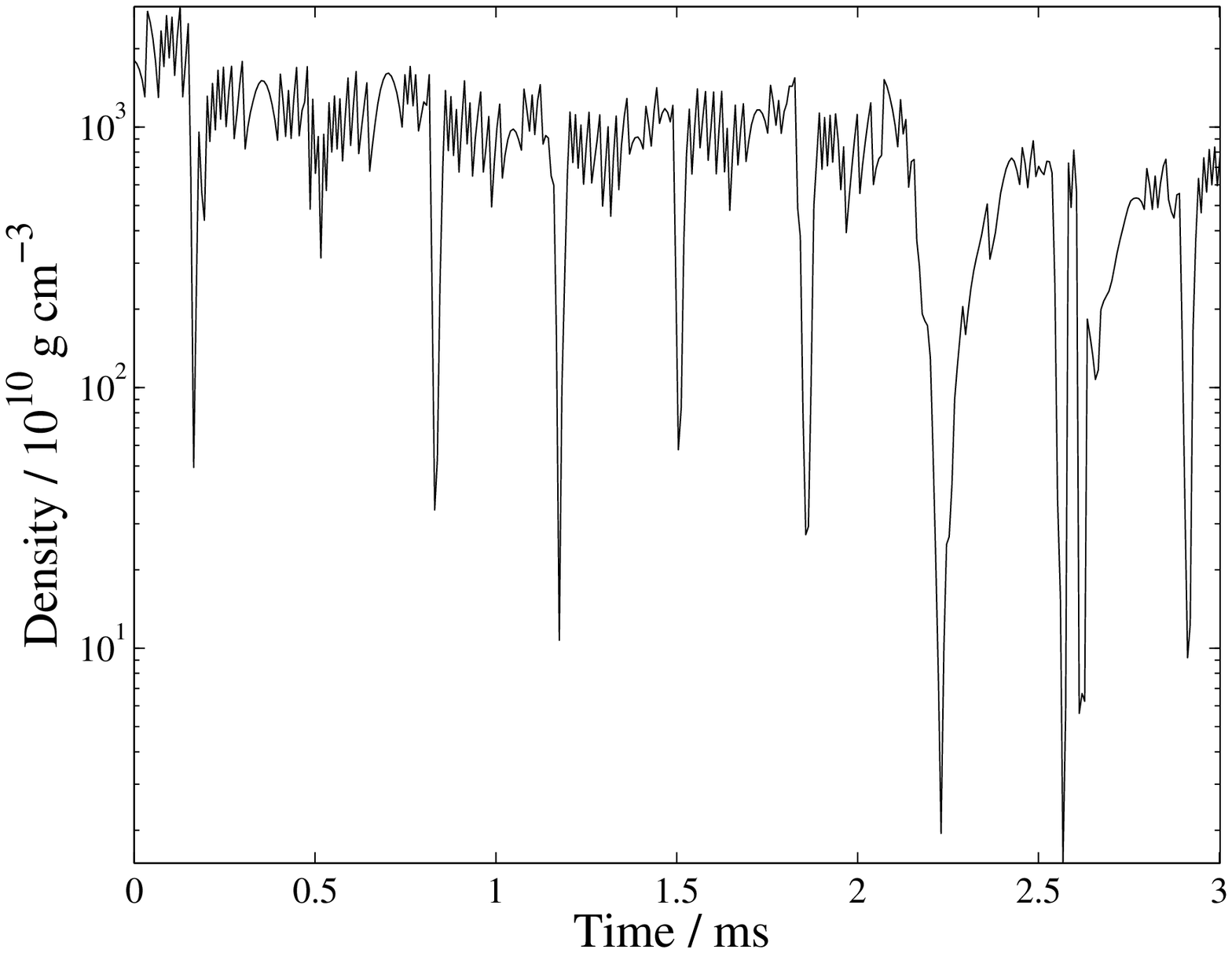}
\hfill\includegraphics[scale=0.4,keepaspectratio = false]{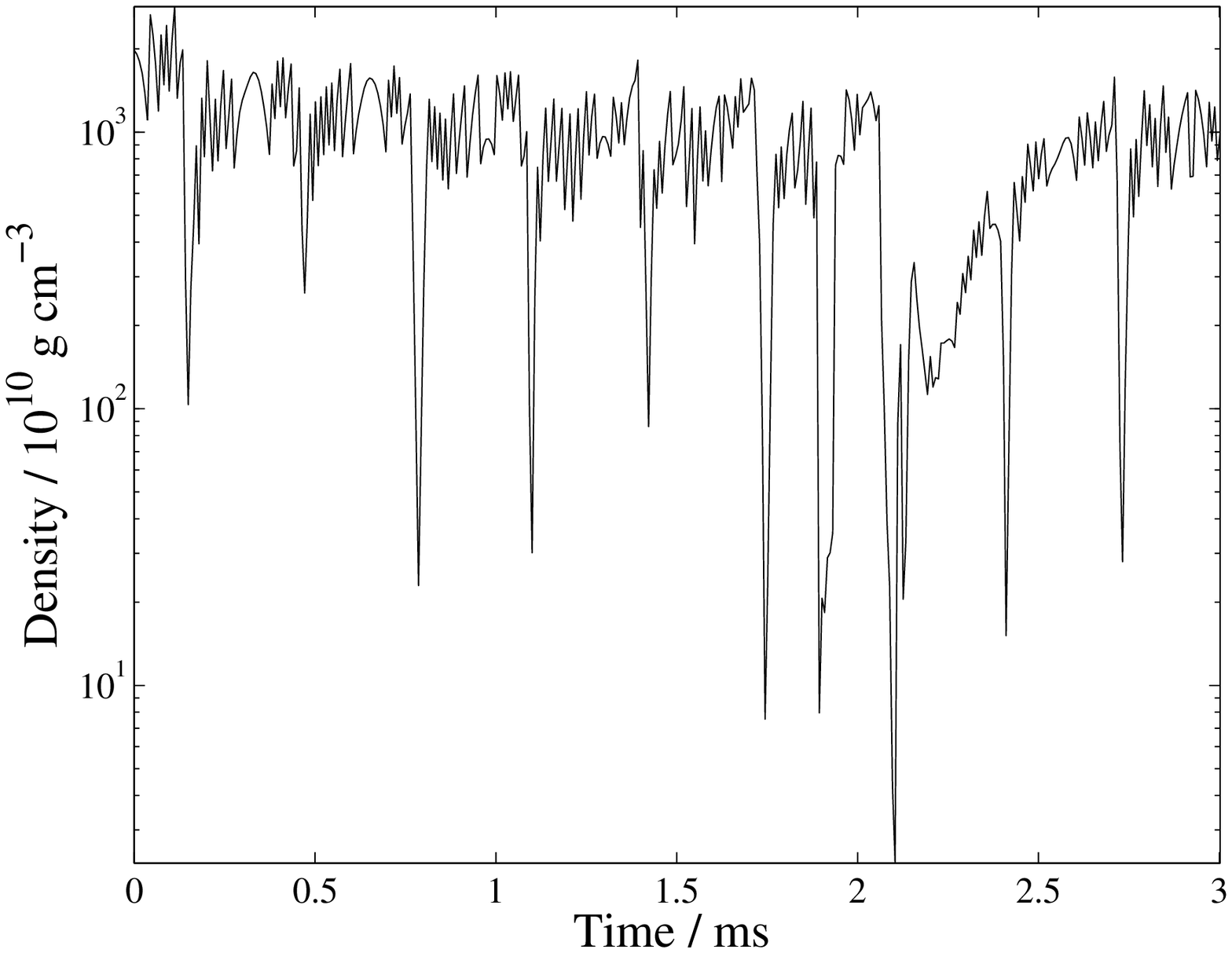}
\hfill\includegraphics[scale=0.4,keepaspectratio = false]{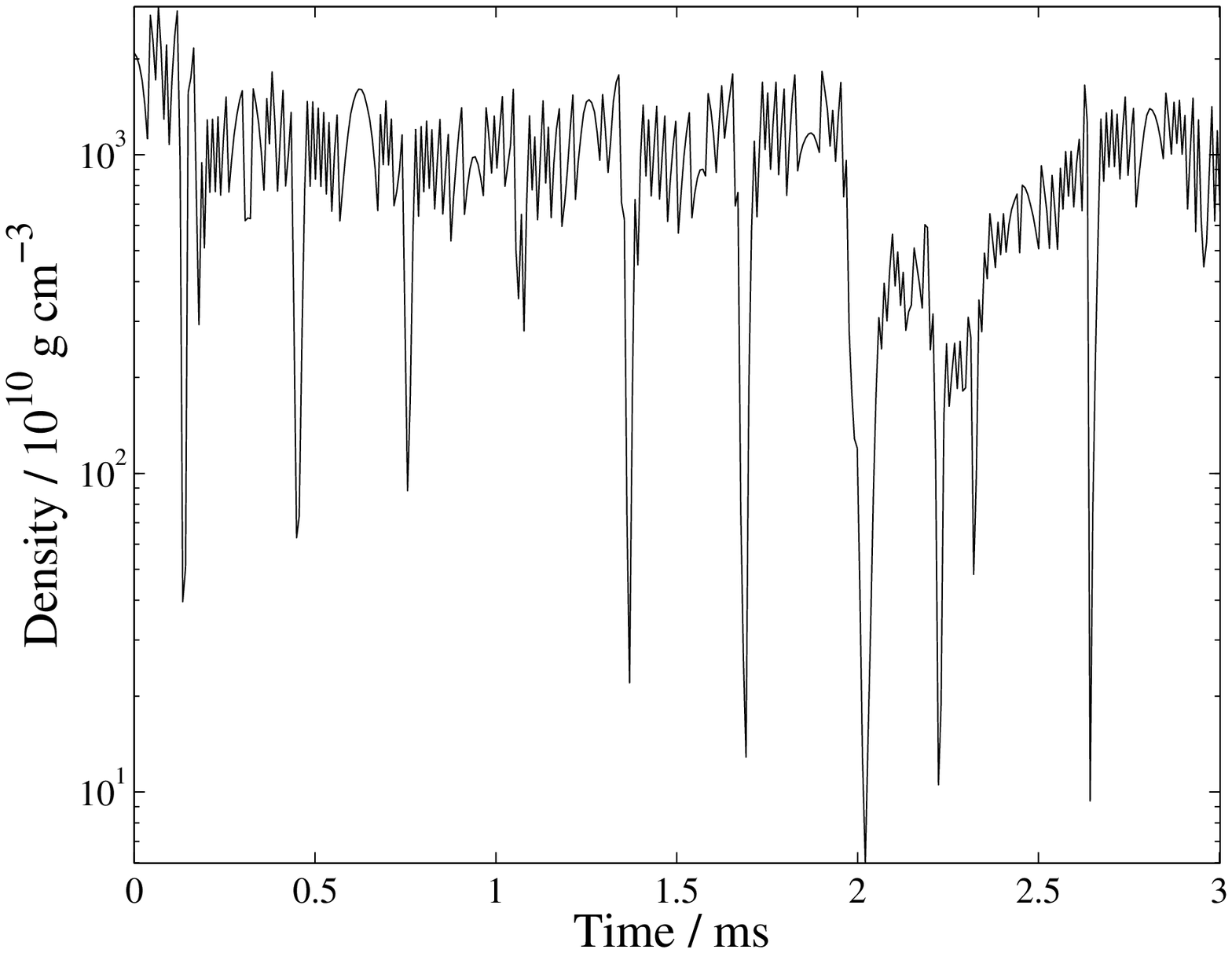}
\hfill\includegraphics[scale=0.4,keepaspectratio = false]{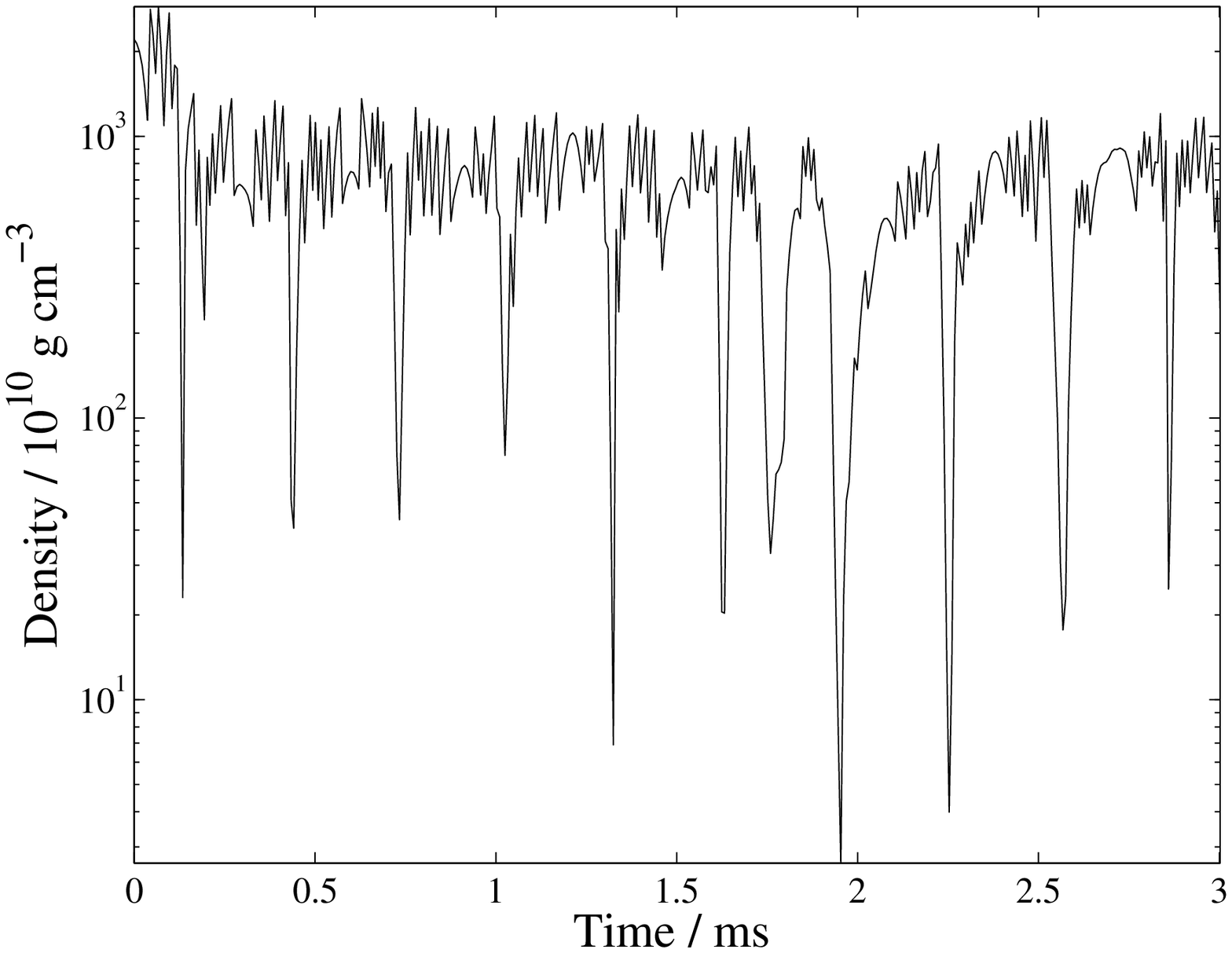}
\hfill\caption{Density at the neutrinosphere versus time for $M=1.55M_{\odot}$, (left upper figure), $M=1.7M_{\odot}$, (right upper figure), $M=1.8M_{\odot}$, (left lower figure), $M=1.9M_{\odot}$, (right lower figure).}
\label{fig:rhonu_time_models}
\end{figure}

With the simulated density and temperature profiles on a given time
slice, we can calculate the position of the neutrinosphere $R_\nu$
(see Eq.~(\ref{eq:Janka2001eq16})) with a trial-and-error method.
Since $R_\nu$ is a function of both temperature and density (cf.
previous  Section), it also oscillates with the same period as the
central density. Figs.~\ref{fig:Tnu_time_models} and
Figs.~\ref{fig:rhonu_time_models} show the temperatures and the densities at
the neutrinosphere as a function
of time for neutron stars with masses of 1.55, 1.7, 1.8 and 1.9
$M_{\odot}$, respectively.

\begin{figure}[ht]
\includegraphics[scale=0.4,  keepaspectratio = false,clip=true ]{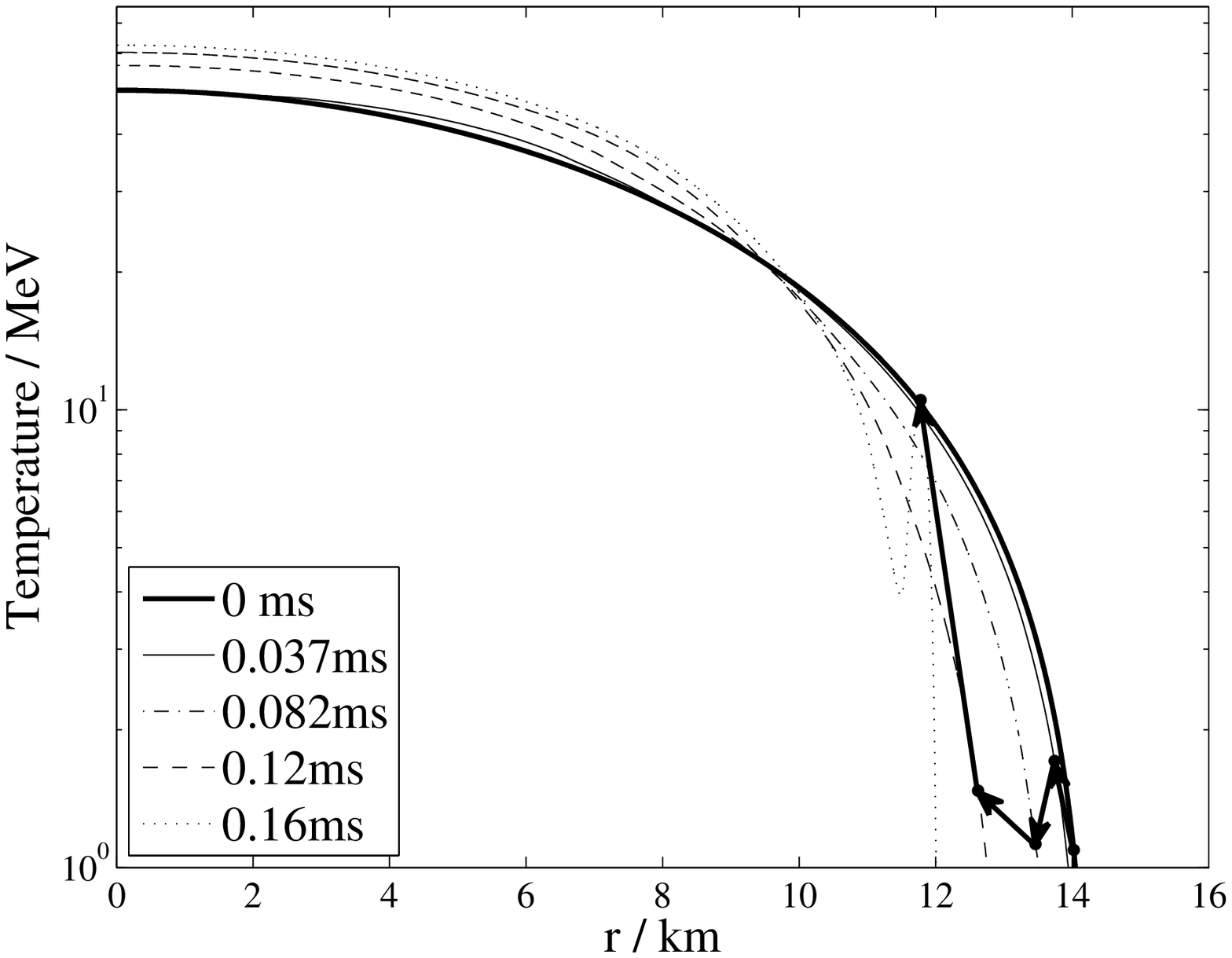}
\hfill\includegraphics[scale=0.4, keepaspectratio = false,clip=true]{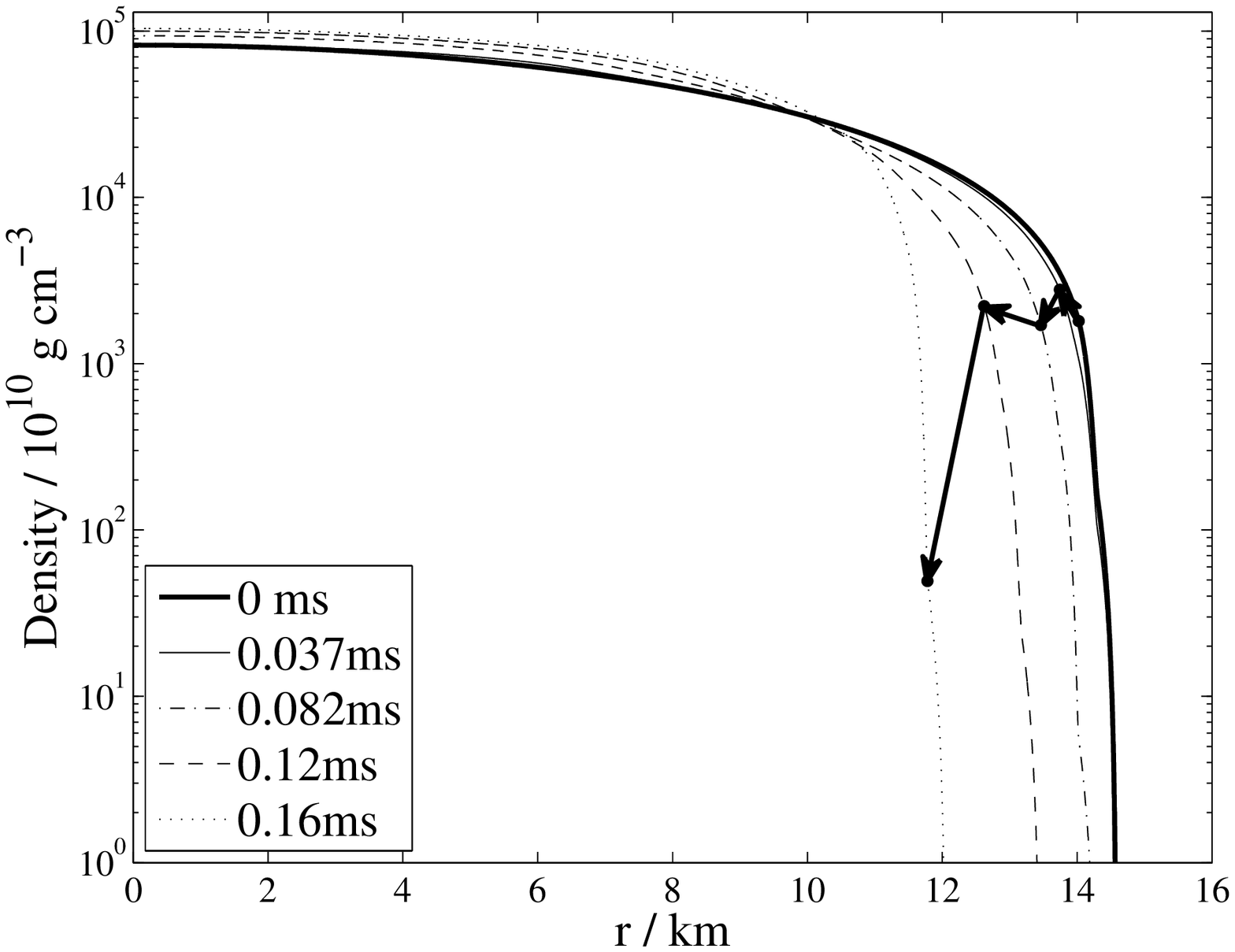}
\caption{Temperature evolution (figure on the left) and density
evolution (figure on the right). The arrow indicates the position at
which the temperature (density) change occurs at the
neutrinosphere.}
\label{fig:Trho_profile_M1.55}
\end{figure}

We can see that both the temperature and
the density at $R_\nu$ are pulsating, with the same period as that
of the central density, but they are almost 180$^{\circ}$ out of
phase to each other.

To show the origin of the pulse like temperature and density
evolution, we pick 5 time slices from $t=0$ to $t=0.16$ ms, and we
focus on the temperature and density evolution at the
neutrinosphere. Figs.~\ref{fig:Trho_profile_M1.55} show the time evolution
of the
temperature and of the density profile of a star with
1.55$M_{\odot}$ respectively.

\begin{figure}[ht]
\includegraphics[scale=0.4,  keepaspectratio = false,clip=true ]{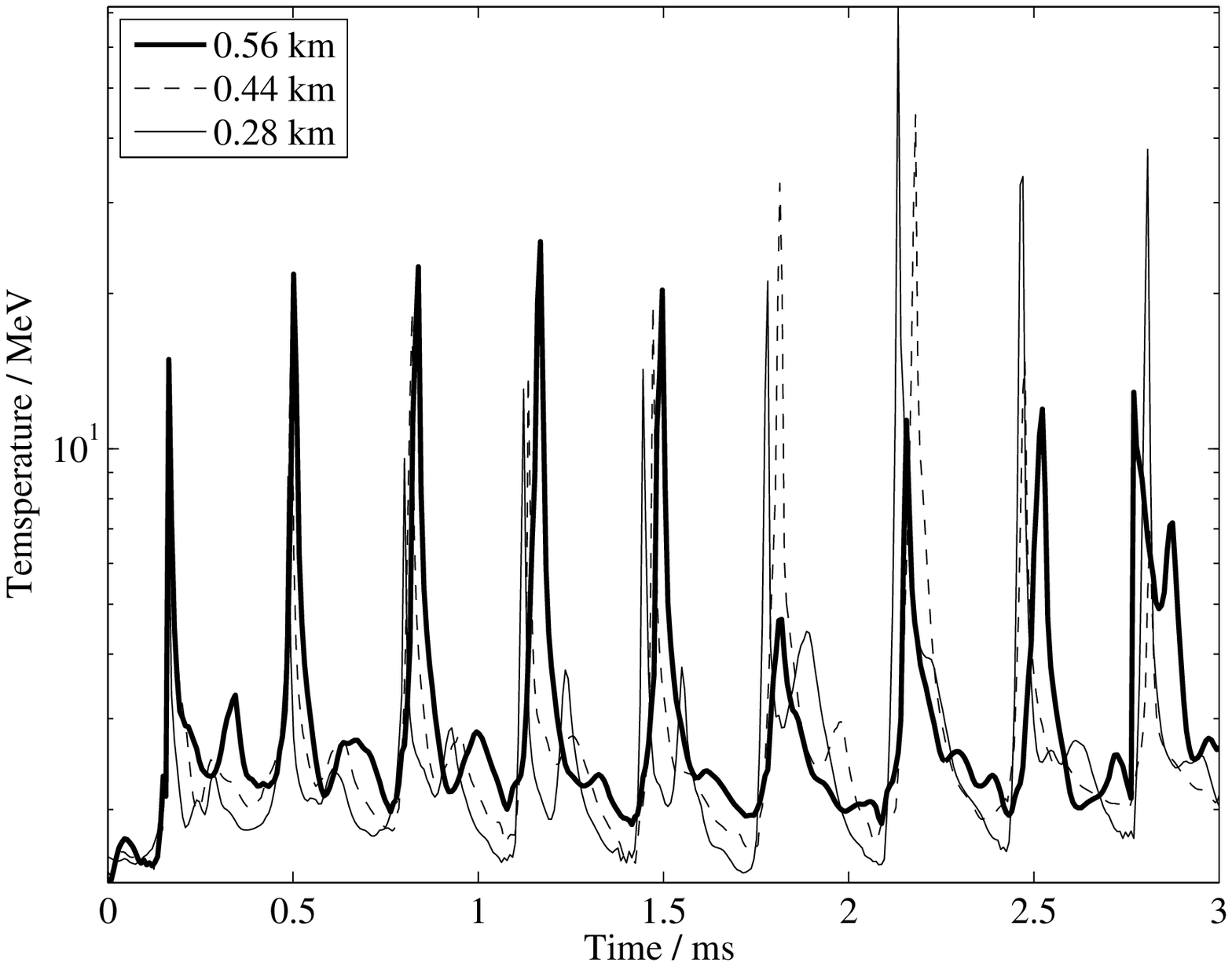}
\hfill\includegraphics[scale=0.4, keepaspectratio = false,clip=true]{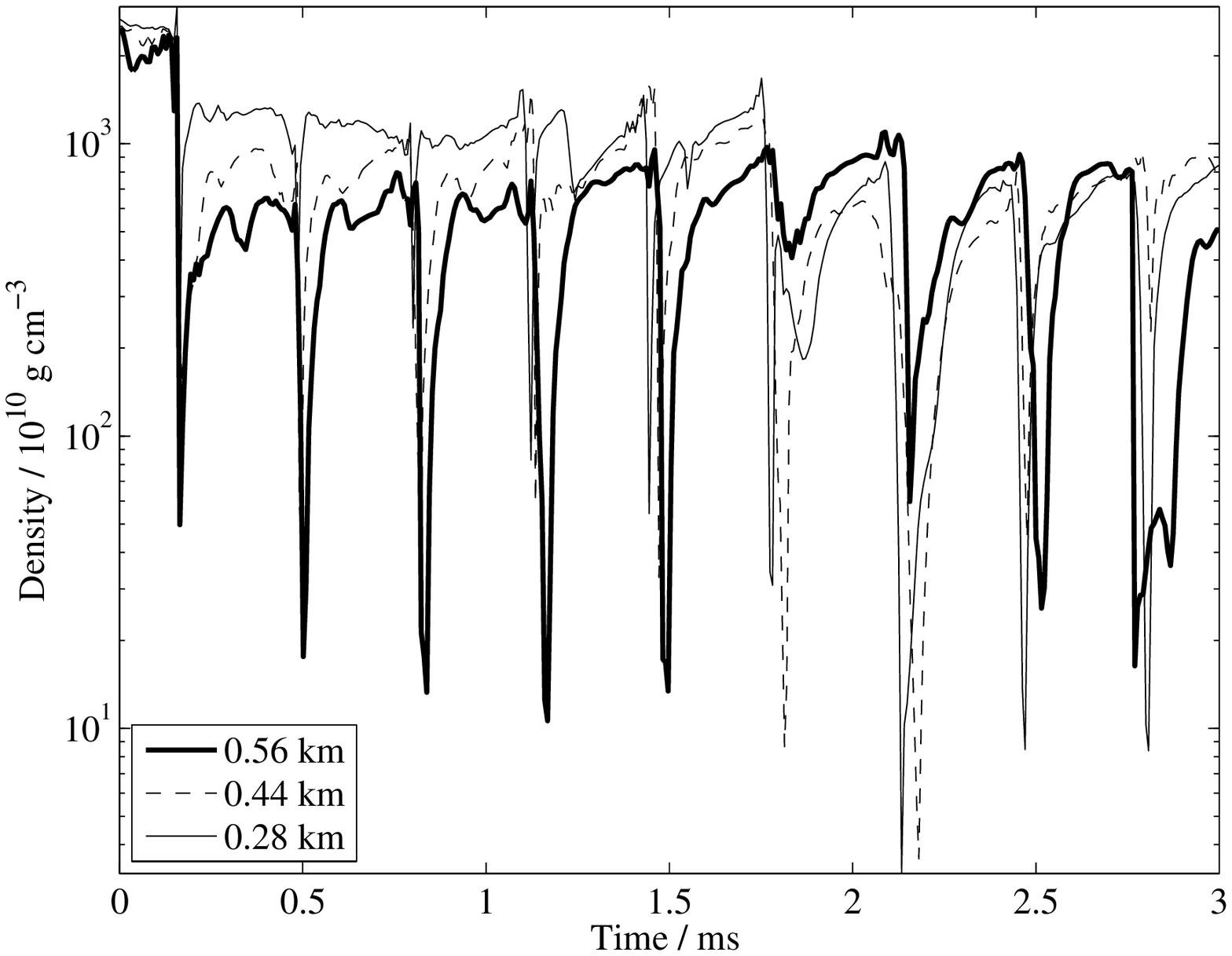}
\caption{Evolutions of the temperature (figure on the left) and density
(figure on the right) at the neutrinosphere for $M=1.75M_{\odot}$, with three
different grid resolutions.
}
\label{fig:grid_Tem_Den}
\end{figure}

We find that the core temperature is
rising immediately after the phase transition, which starts at
$t=0$, and the heat is moving outward. The neutrinosphere is moving
inward because the matter is falling in. The star is shrinking until
$t\sim 0.16$ ms, which is half of the oscillation period, and the
density at the neutrinosphere is also at its minimum. Also at $t\sim
0.16$ ms, the outward heat pulse meets the infalling material
density minimum. This can also be understood from the definition of
the radius of the neutrinosphere, which is defined as the optical
depth equal to unity. Therefore, when the temperature at the
neutrinosphere is maximum, the corresponding density must be at the
minimum value.

While the resolutions we used in the simulations (which are limited
by the computational resource) are good enough to model the global dynamics
of the star accurately, we note that this is not the case near the stellar
surface where the density is very small and changing rapidly.
In particular, the grid resolution near the neutrinosphere is not very high.
We have compared the numerical results obtained with a few different
resolutions in order to examine the effects of resolution.

In Figs.~\ref{fig:grid_Tem_Den}, we show the evolutions of the temperature
and density at the neutrinosphere for the collapse model with $M=1.75M_{\odot}$.
The figures show that the different resolution results agree quite well
qualitatively during about the first 1.5 ms.
In particular, the period of the pulses does not depend strongly on the
resolution. The maximum variation in period for different grid size is given
by $0.28km/C_s$, where $C_s\sim 10^9cm/s$ is sound speed near the surface, and it
gives the maximum shift in period $\sim 0.03ms$.

\section{Emission of neutrinos and $e^{\pm}$ pairs}

\subsection{Neutrino Luminosity}

The neutrino luminosity is given by \cite{Balantekin2005},
\begin{equation}
L_\nu = 4 \pi r^2 c \, \frac{1}{2 \pi \hbar ^2} \int \, \frac{E_\nu \, d^3 \mathbf{p}_\nu}{1+\exp ( E_\nu-\mu_\nu/kT_\nu)}, 
\end{equation}
where $\mu _\nu$ is the neutrino chemical potential. Taking $\mu
_\nu=0$, the neutrino luminosity emitted from the neutrinosphere is
given by $L_\nu= 7\pi R_\nu^2 ac T_\nu^4/16$, where $a=4\sigma /c$
is the radiation constant, $\sigma $ is the Stefan-Boltzmann
constant, and $T_\nu$ is the temperature of the neutrinosphere. If
we assume equal luminosities for neutrinos and antineutrinos, the
combined luminosity for a single neutrino flavor is
\begin{equation}\label{eq:lum}
L_{\nu, \, \overline{\nu}} = L_\nu + L_{\overline{\nu}}= \frac{7}{8}
\pi R_\nu^2 ac T_\nu^4.
\end{equation}

The effect of coherent forward scattering must be taken into account when considering the oscillations of neutrinos traveling through matter \citet{Wo78}. Although different flavor neutrinos have different $R_\nu$, yet they have approximately the same value
of luminosity for all flavors \cite{Janka1995,Janka2001} (for general reviews and in depth discussions of the present status of neutrino oscillations and their astrophysical implications  see \cite{Mal04, Me07, Ak08, Al09}). Therefore the total luminosity is
around three times of a single neutrino flavor luminosity
\begin{eqnarray}\label{eq:lumAll}
L &=& L_{\nu_e,\,\overline{\nu}_e} +
L_{\nu_\mu,\,\overline{\nu}_\mu} +
L_{\nu_\tau,\,\overline{\nu}_\tau}\nonumber\\
&=& \frac{21}{8} \pi R_\nu^{2} ac T_\nu^4.
\end{eqnarray}

\begin{figure}[ht]
\includegraphics[scale=0.4, keepaspectratio = false ]{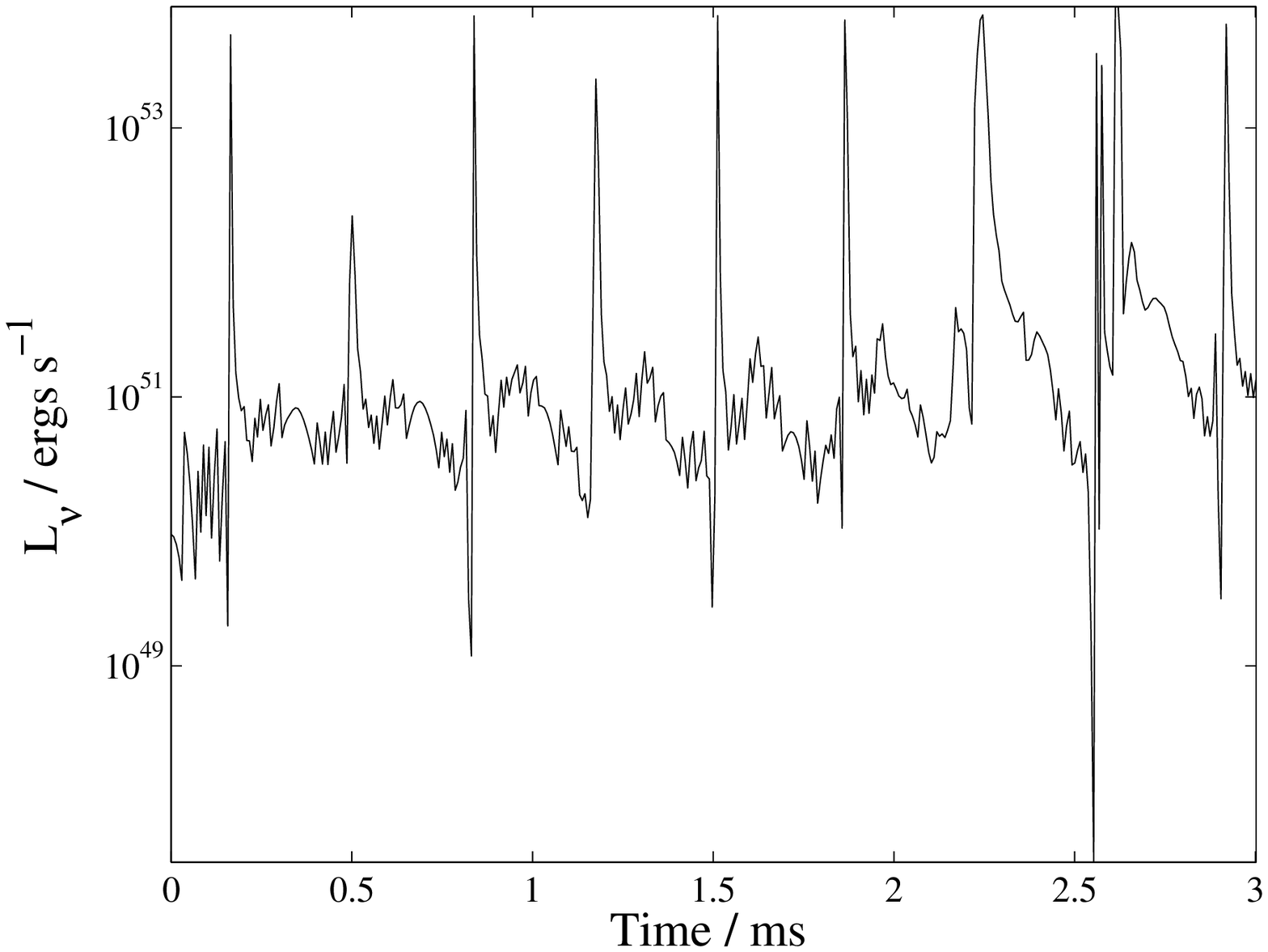}
\hfill\includegraphics[scale=0.4,keepaspectratio = false]{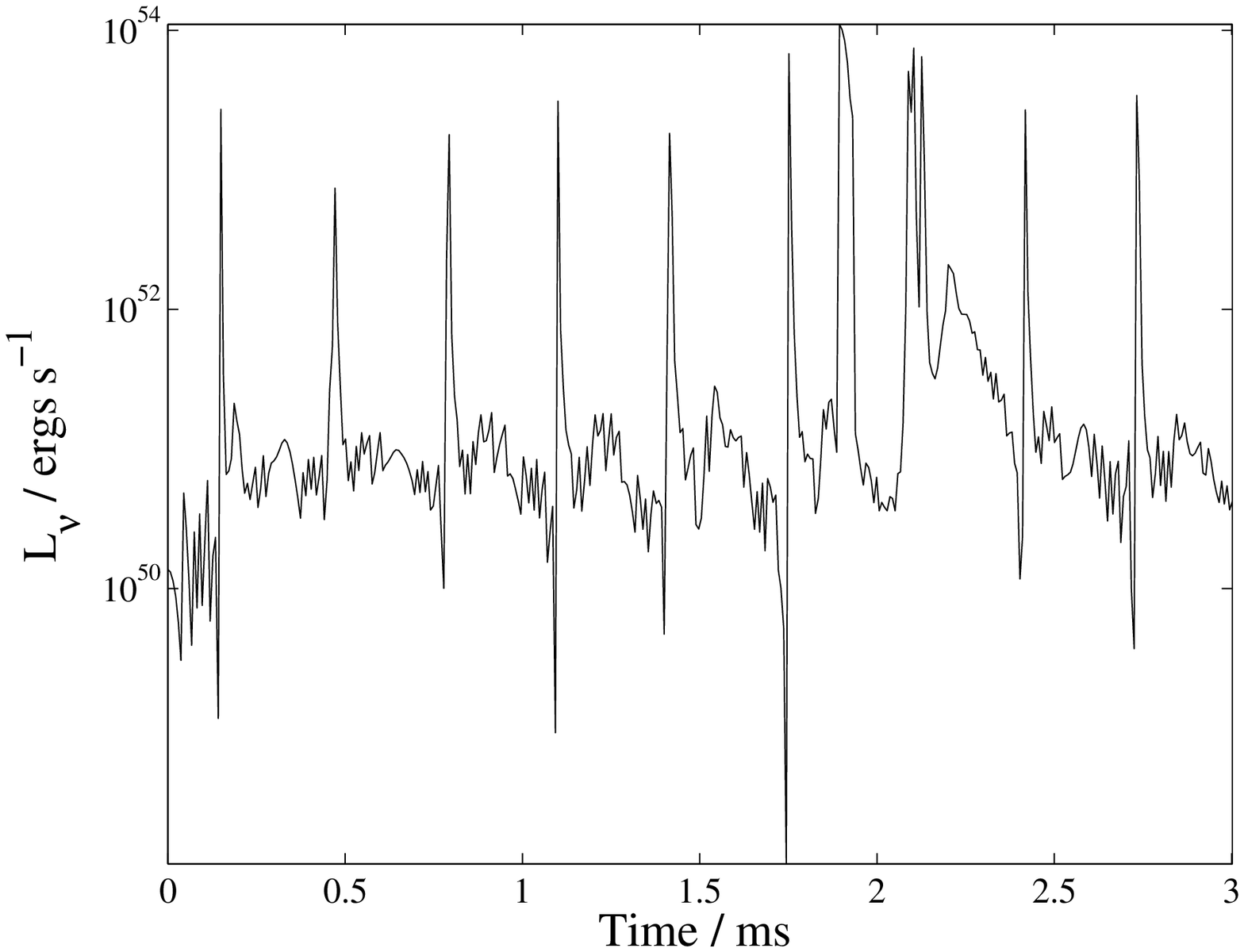}
\hfill\includegraphics[scale=0.4,keepaspectratio = false]{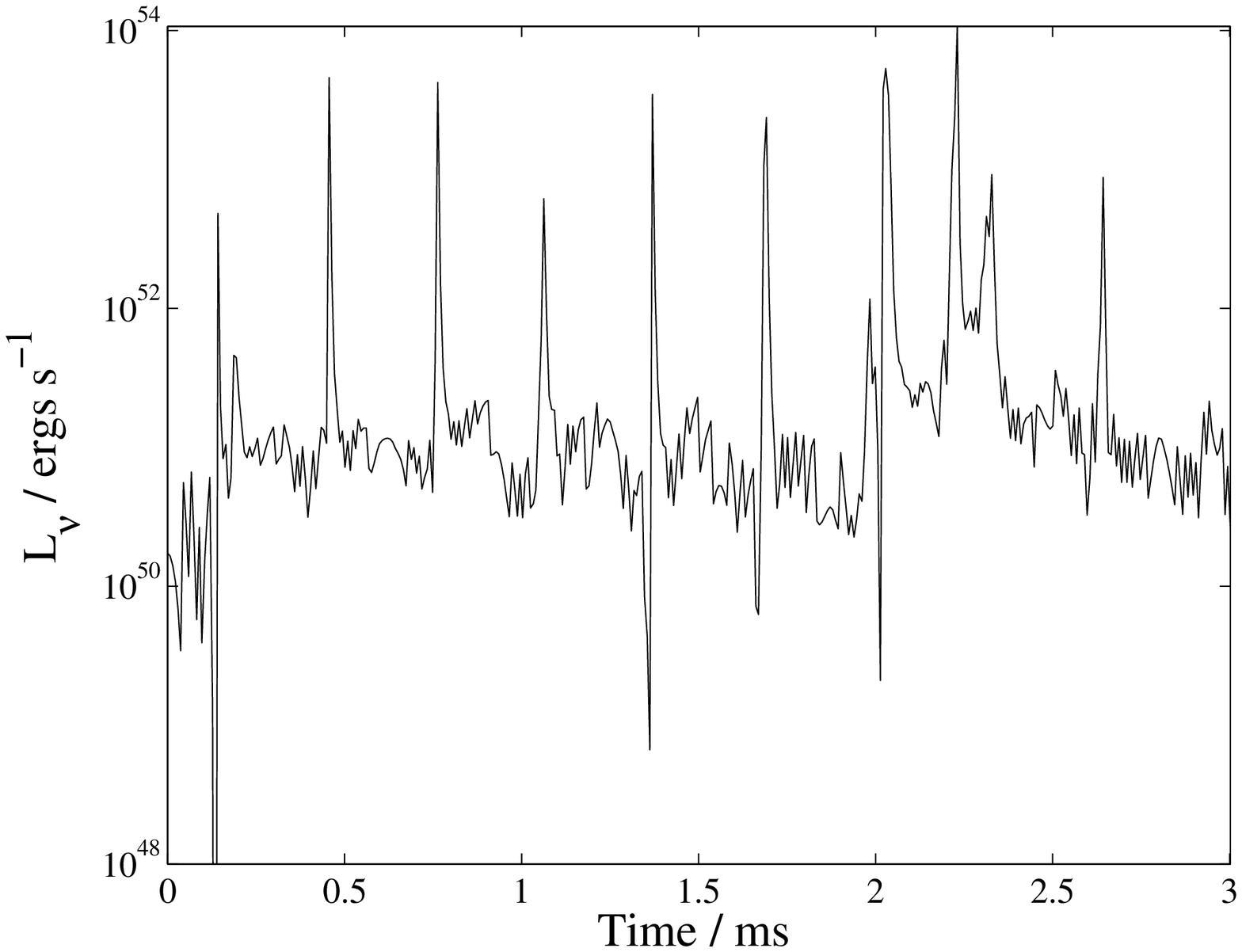}
\hfill\includegraphics[scale=0.4,keepaspectratio = false]{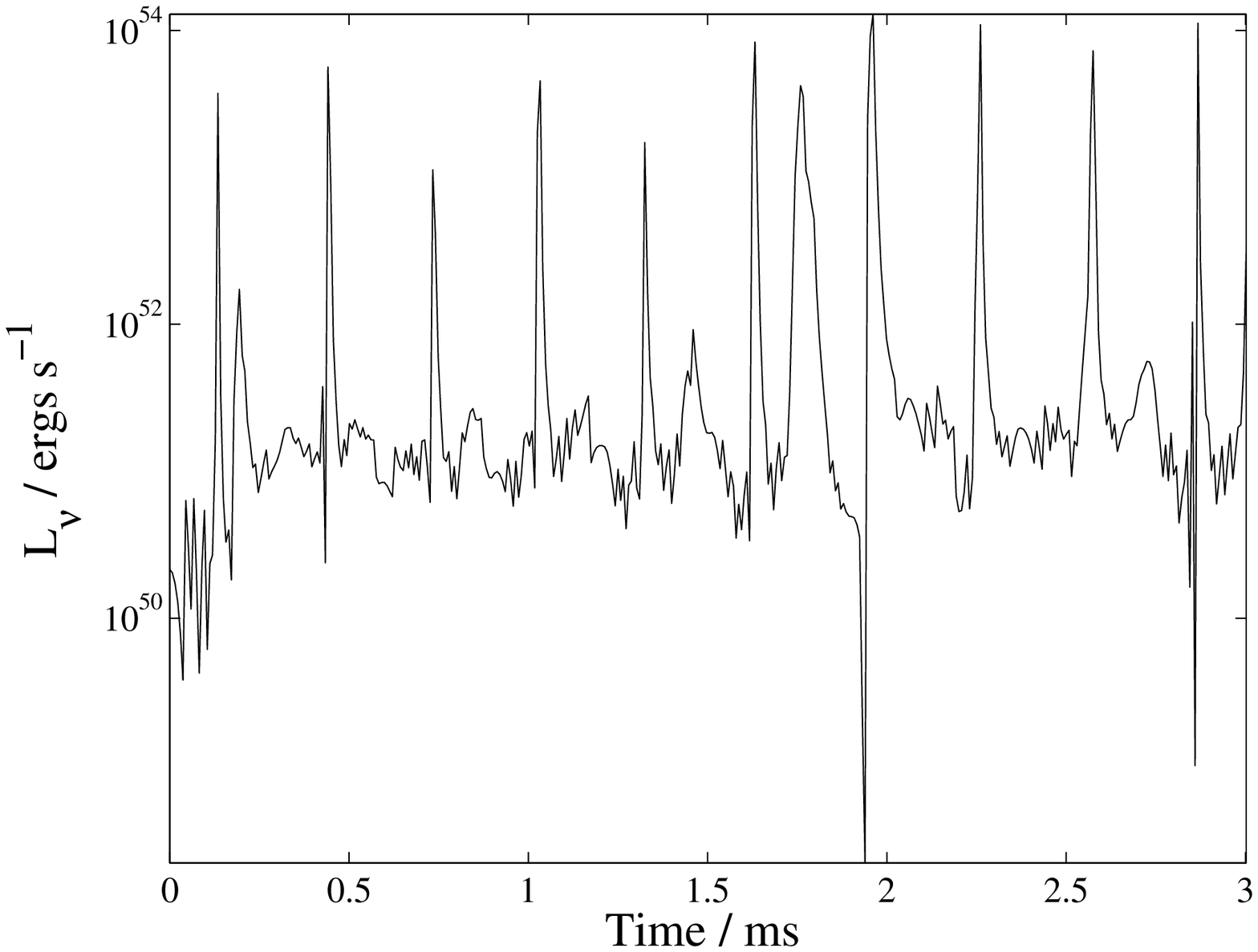}
\caption{Neutrino luminosity versus time for $M=1.55M_{\odot}$, (left upper figure), $M=1.7M_{\odot}$, (right upper figure), $M=1.8M_{\odot}$, (left lower figure), $M=1.9M_{\odot}$, (right lower figure).}
\label{fig:Lnu_time_models}
\end{figure}

Using $R_\nu$ and $T_\nu$ obtained from the last  Section, we
compute the neutrino luminosity as a function of time. The
results are shown in Fig.~\ref{fig:Lnu_time_models}.

The peak luminosities range
from $10^{52}$ to $10^{54}$ ergs/s; the pulsating period of the
luminosity is the same as that of the temperature and of the
density.

\subsection{Pair Production Rate}

Neutrinos and antineutrinos can become electron and positron pairs
via the neutrino and antineutrino annihilation process
($\nu~+~\bar{\nu}~\rightarrow ~e^-~+~e^+ $). The total neutrino and
antineutrino annihilation rate can be given as follows \cite{Goodman1987,Cooperstein1987}
\begin{eqnarray}
\dot{Q}_{\nu \bar{\nu} \rightarrow e^{\pm}}&=&\frac{7 G_F^2 \pi^3
\zeta(5)}{2 c^5 h^6} D \left[kT_\nu(t)\right]^9 \int _{R_{\nu}} ^{\infty}
\Theta(r) \, 4 \pi r^2 dr\\ &=& \frac{7 G_{F}^{2} D \pi^3
\zeta(5)}{2 c^5 h^6} \frac{8\pi^3 }{9} R_{\nu}^3(kT_\nu)^9,
\end{eqnarray}
where $\Theta(r)=2\pi^2(1-x)^4(x^2+4x+5)/3$, $x =
\sqrt{1-R_{\nu}^2/r^2}$, $T_\nu(t)$ is the temperature at the
neutrinosphere at time $t$, $G_F^2=5.29\times 10^{-44}$ is the Fermi
constant,  $\zeta $ is the Riemann zeta function, and $D$ is a
numerical value depending on the pair creation processes (e.g.
experimental results indicate that $D_1=1.23$ for
$\nu_e\,\nu_{\bar{e}}$ and $D_2=0.814$ for $\nu_\mu \,
\nu_{\bar{\mu}}$ and $\nu_\tau \, \nu_{\bar{\tau}}$). To obtain the
total neutrino annihilation rate from all species,
$\nu_{e}\,\nu_{\bar{e}}$, $\nu_{\mu}\,\nu_{\bar{\mu}}$ and
$\nu_{\tau}\,\nu_{\bar{\tau}}$, we sum up the energy rate for each
single flavor,
\begin{eqnarray} \label{eq:Me1}
\dot{Q} &=& \dot{Q}_{\nu_e\,\bar{\nu}_e} +
\dot{Q}_{\nu_\mu\,\bar{\nu}_\mu} +
\dot{Q}_{\nu_\tau\,\bar{\nu}_\tau}\nonumber\\
&=& \frac{28 G_F^2 \pi^6 \zeta(5)}{9 c^5 h^6} (D_1 +
2D_2)R_\nu^3(kT_\nu)^9.
\end{eqnarray}

\begin{figure}[ht]
\includegraphics[scale=0.4, keepaspectratio = false ]{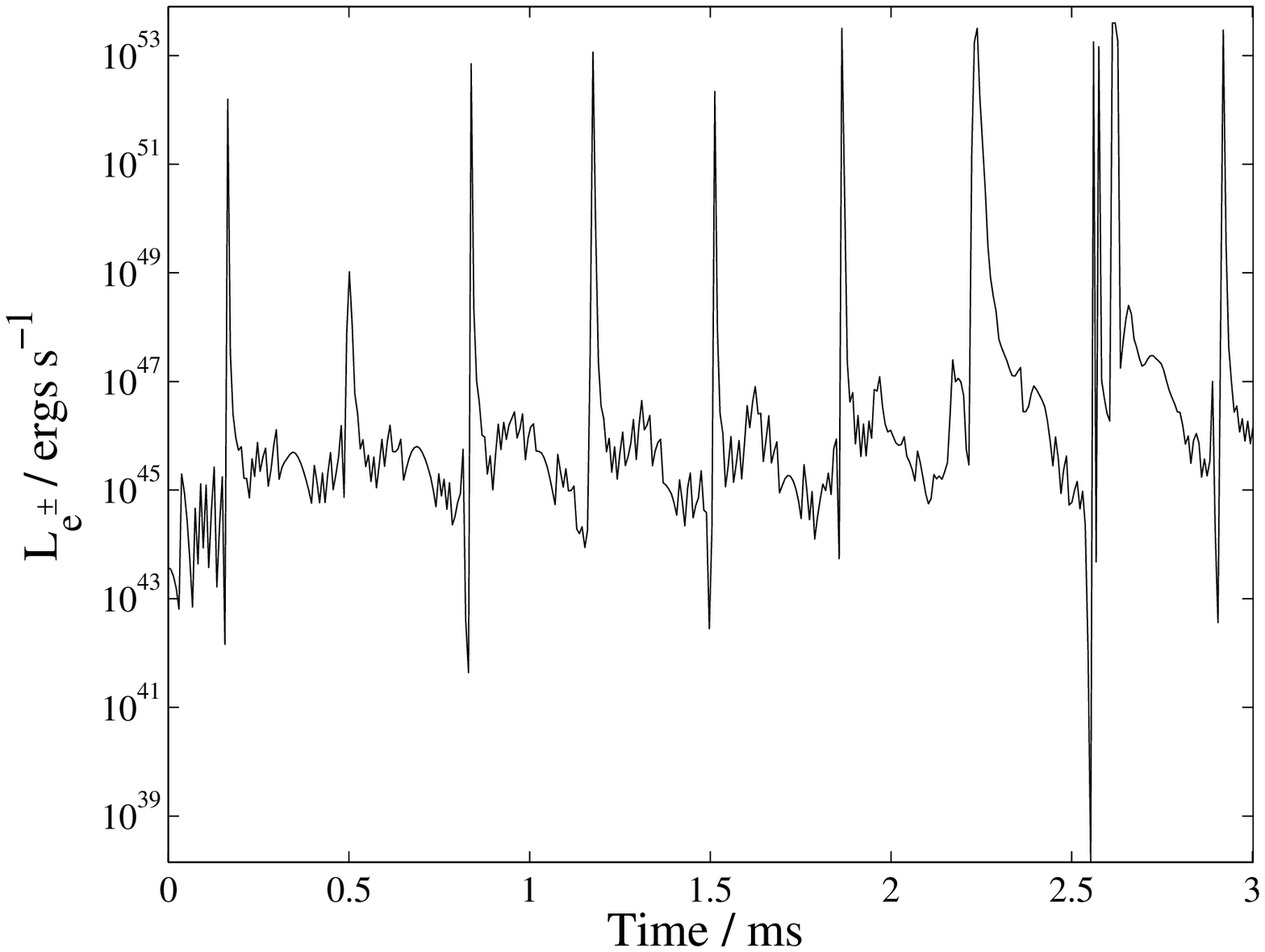}
\hfill\includegraphics[scale=0.4,keepaspectratio = false]{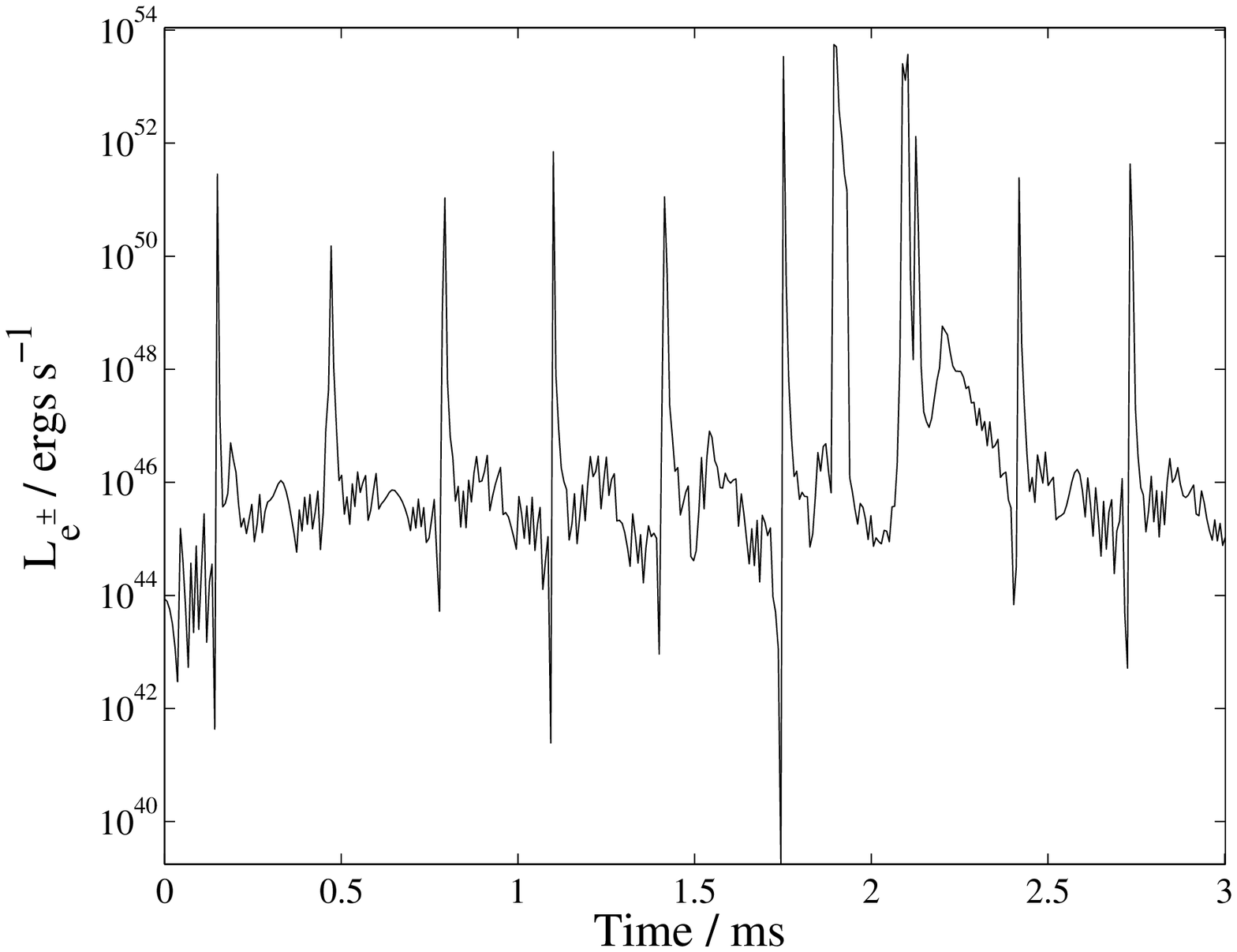}
\hfill\includegraphics[scale=0.4,keepaspectratio = false]{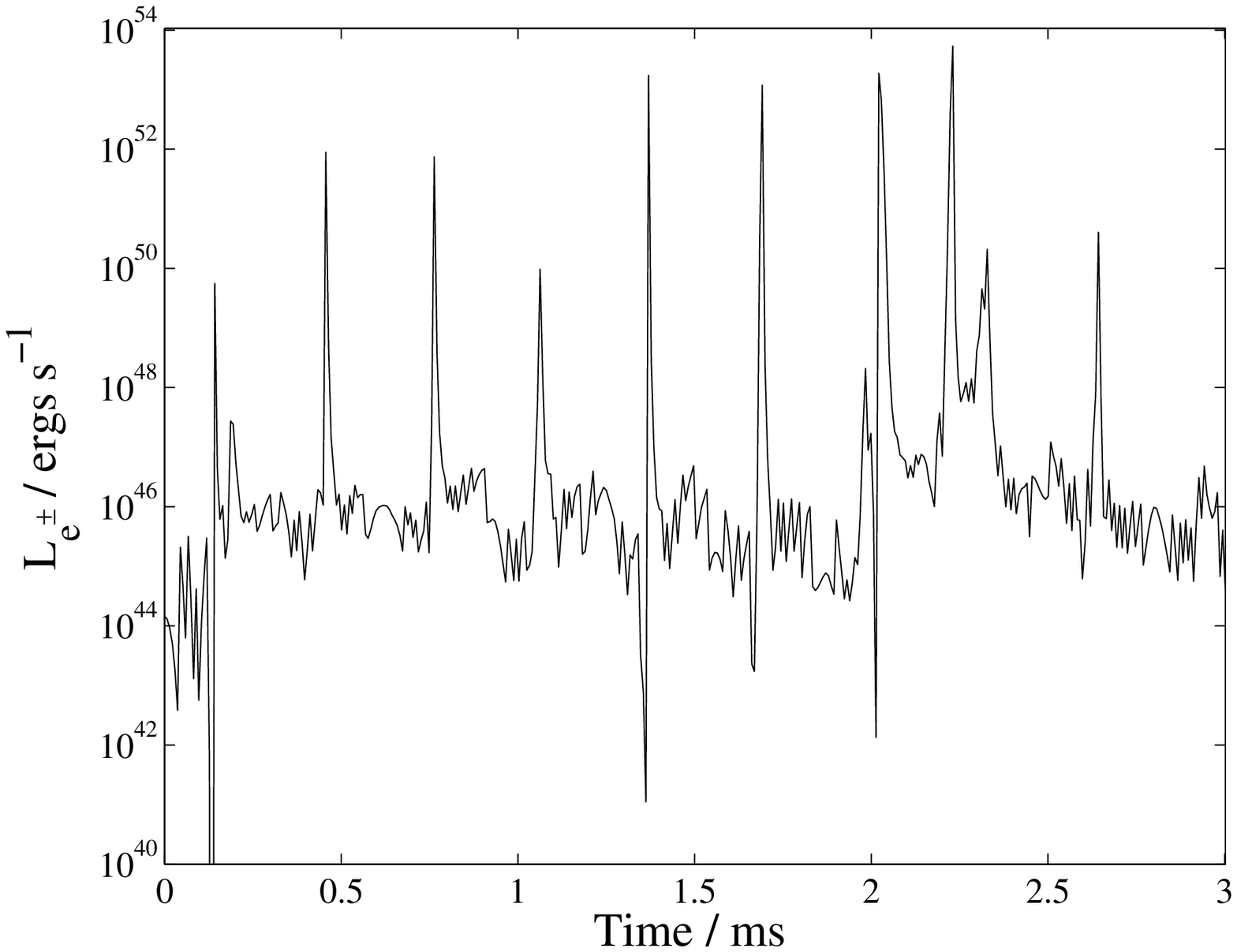}
\hfill\includegraphics[scale=0.4,keepaspectratio = false]{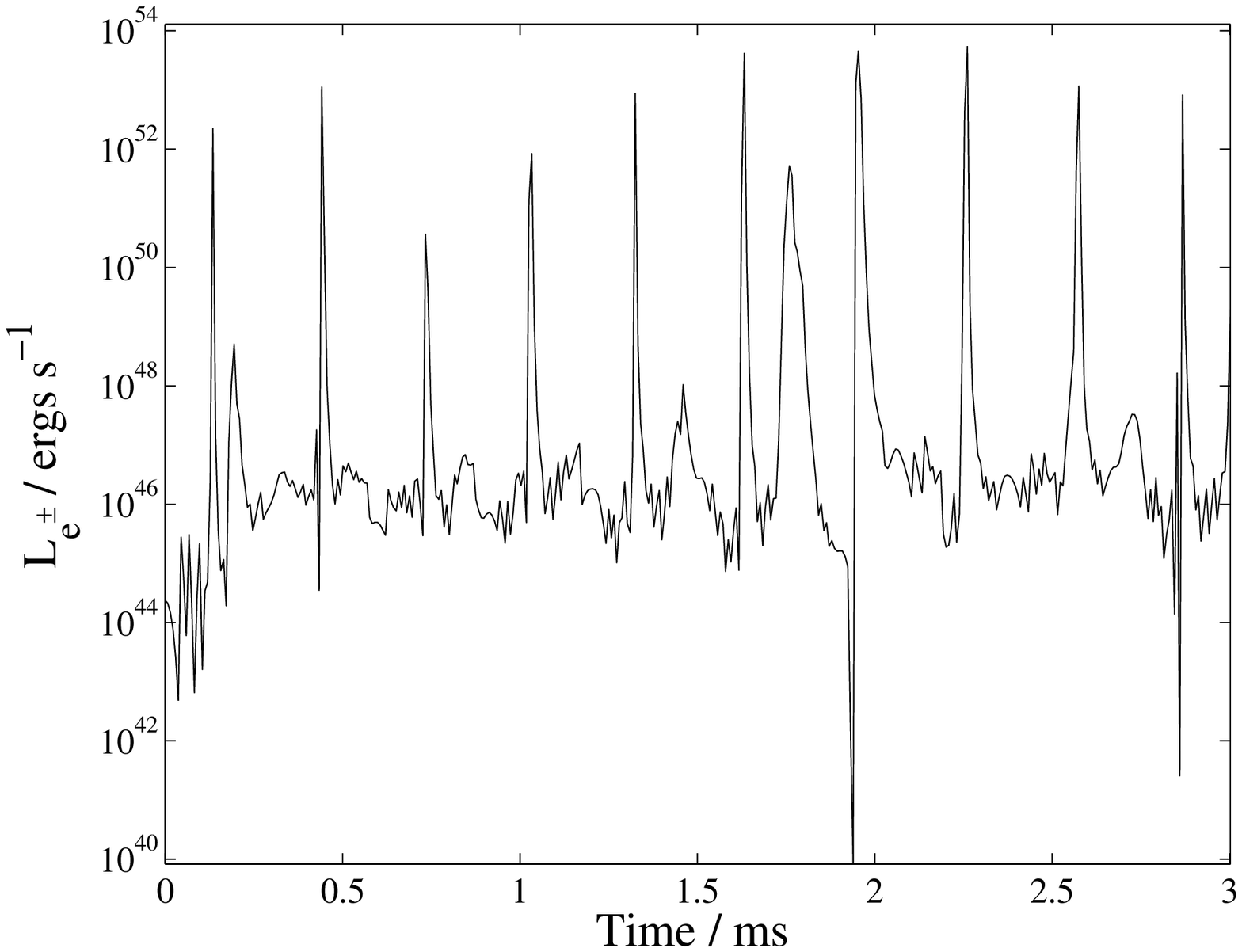}
\caption{Electron-positron luminosity versus time for $M=1.55M_{\odot}$, (left upper figure), $M=1.7M_{\odot}$, (right upper figure), $M=1.8M_{\odot}$, (left lower figure), $M=1.9M_{\odot}$, (right lower figure). }
\label{fig:Le_time_models}
\end{figure}

 Fig.~\ref{fig:Le_time_models} shows the rate of energy carried away by the
electron/positron pairs produced through neutrino annihilation,
which varies from $\sim 10^{51}$ergs/s to $\sim 10^{53}$ergs/s.

It
is interesting to note that almost all neutrinos can be annihilated
into electron-positron pairs at the peak because of the extremely
high density and high energy of the neutrinos.  In particular the
rest mass of the electrons/positrons is much smaller than
$kT_{\nu}$.

\section{Mass ejection and acceleration}

\begin{figure}[ht]
\centering
\includegraphics[scale=0.9, keepaspectratio = false ]{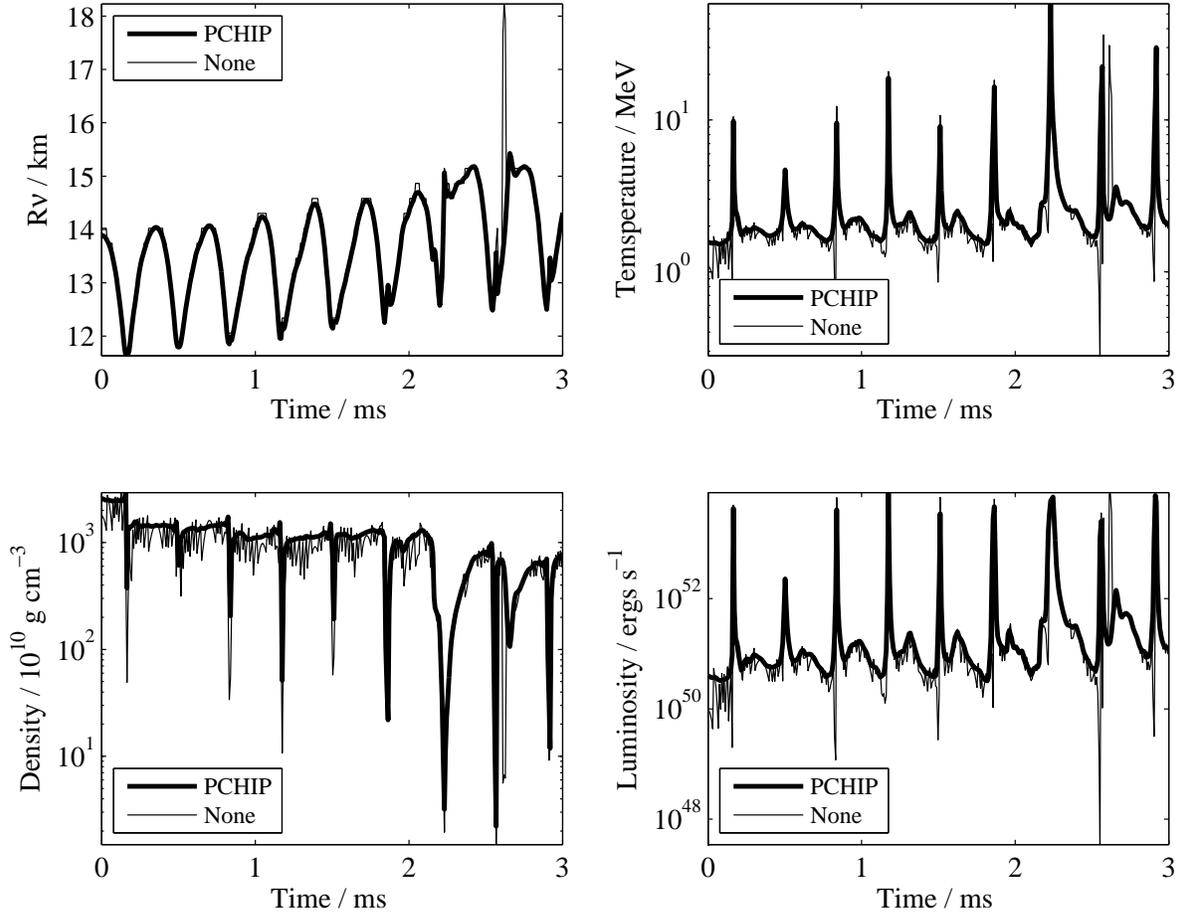}
\caption{Results with PCHIP method and without data interpolation
for M=1.55$M_{\odot}$. }
\label{fig:PCHIP}
\end{figure}

In order to calculate the mass ejected from the stellar surface by
neutrinos/antineutrinos and pairs, a very detailed knowledge of the
mass distribution near the surface is required.
It was pointed out that the density profile in the crust plays
a very important role in determining how much mass can be ejected \cite{K05}. They use a static
star model and an assumed simple power law density profile to demonstrate that
the mass ejection can be significantly different.
In our computer
capability the minimum spacial grid size that can be achieved is
0.28 km. In order to estimate a precise location we choose to use
the Piecewise Cubic Hermite Interpolating Polynomial (PCHIP) method
to interpolate the density and  the temperature data along the
grids. The PCHIP method can provide a
more accurate representation of the physical reality \cite{FrCa80}. As compared
with other interpolation methods (e.g., cubic spline data
interpolation), the curve produced by the PCHIP method does not
contain extraneous "bumps" or "wiggles", meaning that it could
preserve the shape of the density and of the temperature profile,
even when they change dramatically. However, it is unavoidable that
even if we choose the best possible method, the true location might
be slightly different from the real one. In Fig.~\ref{fig:PCHIP} we
compare the original time evolution of the position of the
neutrinosphere ($R_{\nu}$), temperature at $R_{\nu}$, density at
$R_{\nu}$, and the neutrino luminosity with the results obtained by
using the PCHIP method.

We can see that these quantities are
qualitatively the same. We believe that the real continuous mass
distributions near the stellar surface can be approximated by a
continuous mass distribution function, obtained from the numerical
simulated data by using a PCHIP method.

The mass ejection from a newly born quark stars was calculated in \cite{K05}.
They argue that neutrino-electron scattering is the most dominated process to deposit
the neutrino energy in the crust. In this paper we argue that the dominated energy deposition
process is the neutrino-antineutrino annihilation process. It should be noted that the optical depth
($\tau$), which is defined as $\tau = n \sigma l$, is the most important quantity to determine
the energy deposition instead of cross section ($\sigma$) alone. Here $n$ is the scattered
particle density and $l$ is the characteristic length. We can only eject mass above neutrinosphere,
which is located only several grid sizes below the stellar surface when the neutrino luminosity
is maximum. From Fig.~4 we can see that the density is several $10^{11}$g/cm$^3$ when the
neutrino luminosity is maximum, where most of the mass is ejected. The density has dropped a factor of 10
in one grid size from the neutrinosphere to the stellar surface. Therefore the scale length
of density is about half of grid size, i.e. $l \sim 0.14$km. In a neutron rich matter, we can take electron fraction as
0.2 then we obtain $n_e l \sim 10^{39}$cm$^{-2}$. On the other hand, the neutrino density is uniform from
the neutrinosphere to the surface of the star and the
density of neutrino is given by $n_{\nu}=11 a T^4/4kT\sim 10^{36}\left(kT/15\;{\rm MeV}\right)^3$, where $a$ is the Stefan-Boltzmann constant and $k$
is the Boltzmann constant. At the neutrino
luminosity maximum, $kT\sim$15 MeV and the distance from the neutrinosphere to the stellar surface is several grid sizes,
which is $\sim$1km. Then $n_{\nu}  l \sim 10^{41}$cm$^{-2}$. Since neutrino-antineutrino annihilation cross section
(cf. Eq.~2 in \cite{Af00}) and neutrino-electron scattering
cross section (cf. Eq.~7 in \cite{K05}) are almost the same so we will ignore this process in calculating the mass ejection.

\subsection{Energy deposition in the crust and mass ejection}

Although most of the neutrinos/antineutrinos created above the
neutrinosphere can escape, part of them can still be absorbed in the
crust. We can estimate the amount of neutrino energy $E_{\nu}$
deposited in the crust due to the absorption in the following way.
If we define $R_{\rm M}$ as $R_{\rm NS}
> R_{\rm M}> R_\nu$, then the absorbed neutrino energy onto the
surface mass layer between $R_{\rm M}$ and $R_{\rm NS}$ could be
expressed as
\begin{equation}\label{eq:NE}
E_{\nu}(R_{\rm M}) = \int \left[1 - e^{-\tau(R_{\rm M})}\right] \,
L(R_{\rm M}) \, dt,
\end{equation}
where $\tau(R_{\rm M})=\int_{R_{\rm M}}^\infty dr \,
\kappa_{eff}(r)$ is the optical depth at $R_{\rm M}$, $L(R_{\rm M})=
21\pi R_{\rm M}^2 \, a \, c \, T(R_{\rm M})^4 /8$ is the neutrino
luminosity above $R_{\rm M}$ and $T(R_{\rm M})$ is the temperature
at $R_{\rm M}$. Notice that we only performed data output for every
20 iterations in our simulations, corresponding to a time interval
$\Delta T = 0.0075$ ms. Hence, the time interval $dt$ in the
integral is taken to be $dt = \Delta T$.

The annihilated pairs created in the crust will be absorbed because
of the much stronger interaction matter than that of neutrinos, and
the pair energy ($E_{l^{\pm}}$) deposited in the crust is given by
\begin{equation}\label{eq:QE}
E_{l^\pm} = \int \dot{Q}(R_{\rm M}) \, dt,
\end{equation}
\begin{equation}
\dot{Q}(R_{\rm M})=\frac{7 \, G_F^2 \, \pi^3 \, \zeta(5)}{2 \, c^5 \, h^6} (D_1 + 2D_2) (kT_\nu)^9 \int _{R_{\rm M}} ^{R_{\rm NS}} \Theta(r) \, 4 \pi \, r^2 \, dr. 
\end{equation}
Note that we only integrate $r$ from $R_{\rm M}$ to $R_{\rm NS}$
instead of integrating from $R_\nu$ to $\infty$.

The gravitational binding energy of the surface mass is
\begin{equation} \label{eq:GE}
E_{\rm G} = \frac{G \, M \, \Delta m}{R_{\rm M}} \ ,
\end{equation}
where $M = 4 \, \pi \int _{0}        ^{R_{\rm M}}  r^2 \, \rho(r) \,
dr $,  and $\Delta m (R_{\rm M})= 4 \, \pi \int _{R_{\rm M}}^{R_{\rm
NS}} r^2 \, \rho(r) \, dr$,  respectively.

Since the neutrino and pair absorption inside the neutron star
actually happen simultaneously, we combine the absorbed neutrino and
pair energy together to be $E^{\rm absorbed}$
\begin{equation}
E^{\rm absorbed}=E_{l^\pm}+E_\nu \ .
\end{equation}
As long as $E^{\rm absorbed} (R_{\rm M})> E_{\rm G}(R_{\rm M})$, the
surface layer of $\Delta m (R_{\rm M})$ could be ejected from the
neutron star. Hence from this criteria we can obtain the maximum
ejected mass.

\subsection{Acceleration by pairs}

As we have mentioned in Section 4.2, the neutrino and antineutrino
annihilation is very high at the peak of the neutrino pulses due to
the extremely high density and the high energy of neutrinos. In fact
most pairs are created outside the star, and therefore after the
matter is ejected, it will be accelerated by absorbing the pairs
created by the annihilation processes. The annihilation energy
created from the neutron star surface $R_{NS}$ to $r>R_{NS}$ is
given by $E_{l^{\pm}} = \int \dot{Q}_{\nu \bar{\nu} \rightarrow
e^{\pm}}(r,t) dt$, where
\begin{equation}
\dot{Q}(r,t)=\frac{7 G_F^2 \pi^3 \zeta(5)}{2 c^5 h^6}
(D_1 + 2D_2) \left[kT_\nu(t)\right]^9 \int _{R_{NS}} ^{r} \Theta(r') \, 4 \pi
r'^2 dr'.
\end{equation}

In the following we will briefly describe how the pairs
accelerate the ejected matter. Before we present our calculations,
we first describe the continuous mass ejection processes. The time
slice interval of our output data is $\Delta T = 0.0075$ ms. We can
calculate the maximum amount of ejected mass only time slice by time
slice.

\begin{figure}[ht]
\centering
\includegraphics[scale=0.75, keepaspectratio = false ]{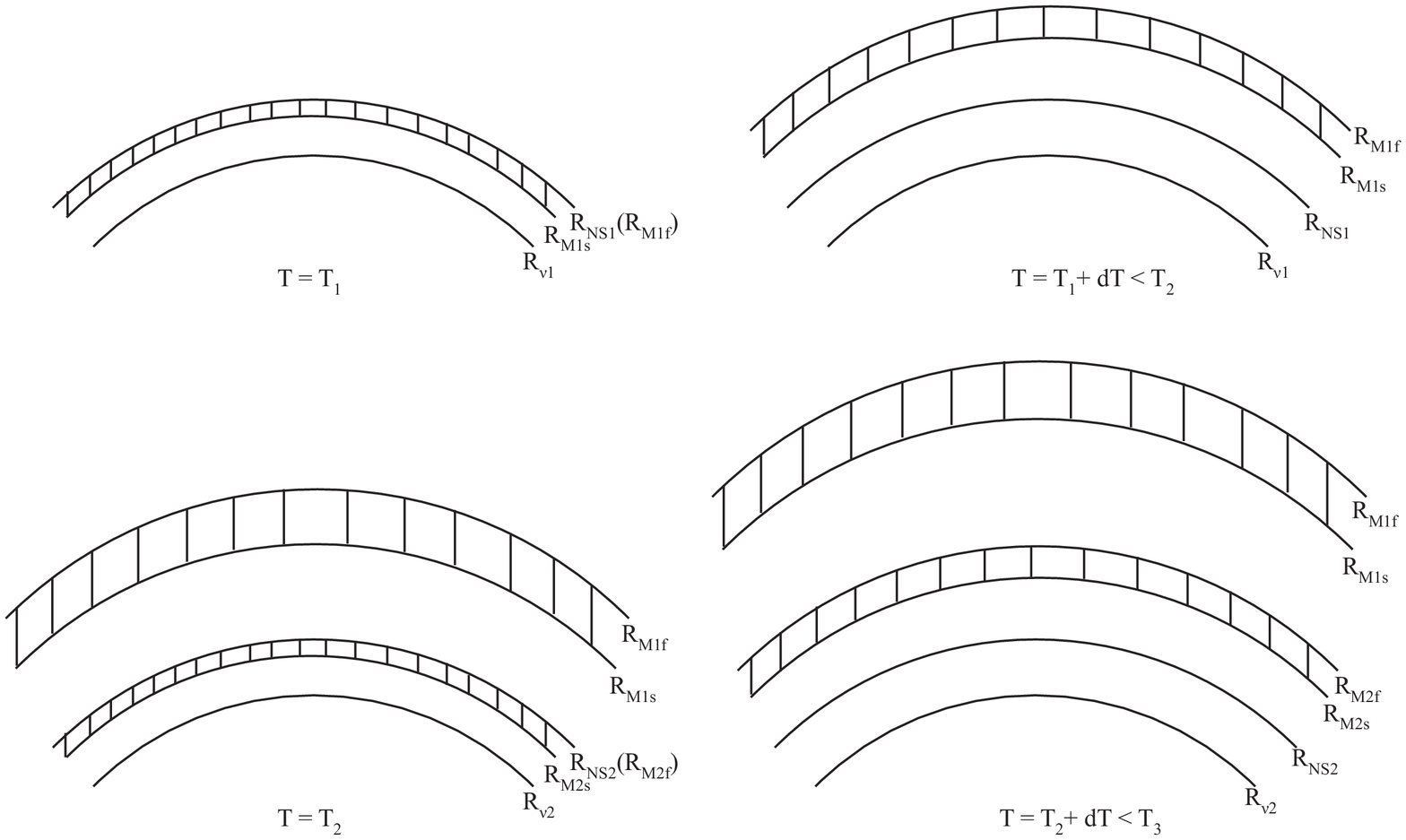}
\caption{Schematic illustration of mass ejection from the stellar
surface.}
\label{fig:mass_eject}
\end{figure}

At $T_1$ (Fig.~\ref{fig:mass_eject} upper left), a layer of mass $\Delta
M(R_M)$ is ejected when $E^{\rm absorbed}(R_M)
> E_{\rm G}(R_M)$. The outer surface $R_{M1f}$ of the ejected mass
is approximately $R_{NS}$, and the velocity of the mass at the outer
surface is almost $c$; the inner surface of the ejected mass is
$R_{M1s}$, where the velocity of the mass at the inner surface is
the escaping velocity, which is almost half of the speed of light.

$T_2$ is the next time step when another layer of mass could be
ejected. Before $T_2$ (Fig.~\ref{fig:mass_eject} upper right), when the mass
layer ejected at $T_1$ is flying outwards from the star at $T_1 + t<
T_2$, the distance to the ejecta can be approximated as $R_{NS1}+ct$
and $t$ is the elapsing time counted from $T_1$. In the space
between $R_{NS1}$ and $R_{NS1}+ct$, pairs are created and they can
move faster than the ejected mass layer. Eventually they are
absorbed by the ejecta and accelerates the mass layer in front.

Another mass layer is ejected at $T_2$ (Fig.~\ref{fig:mass_eject} lower left).
When this ejecta moves outward, it will absorb pairs created between
the stellar surface and this ejecta. However pairs created in the
space between these two ejecta can still be absorbed by the first
ejecta (cf.~Fig.~\ref{fig:mass_eject} lower right). The total pair energy for
accelerating the mass layer ejected at $T_1$ is
\begin{equation}
E^{pair} = \displaystyle \sum_{T_i = T_1}^{T_\infty} \int _{T_i}
^{T_i+\Delta T} \dot{Q}_{T_i}(r_1, r_2)dt,
\end{equation}
where
\begin{equation}
\dot{Q}_{T_i} (r_1, r_2)=\frac{7 G_F^2 \pi^3 \zeta(5)}{2 c^5
h^6} (D_1 + 2D_2) (kT_\nu)^9 \int _{r_1} ^{r_2} \Theta(r) \, 4 \pi
r^2 dr,
\end{equation}
is always calculated between the stellar surface and the ejecta or
between two ejecta.

The Lorentz factor of each ejecta can be estimated as
\begin{equation} \label{eq:gama}
\Gamma = \frac{E^{pair} + E^{absorbed}}{m c^2} + 1.
\end{equation}
We summarize the results of the ejected masses in Tables 1-4. We can
see that sometimes more than one layer of mass can be ejected in one
pulse.

\begin{table*}
 \centering
 \begin{minipage}{140mm}

  \begin{tabular}{|c|c|c|c|c|}
  \hline
Pulse & $\Gamma $ & Energy(ergs) & Mass (g) & Time (s) \\
 \hline
1      &    17          & $2.2\times 10^{46}$ &$ 1.5\times 10^{24}$ & 0.16\\
2      &$6.4\times 10^4  $& $6.5\times 10^{45}$ &$ 1.1\times 10^{20}$ & 0.50\\
3      &$    1.6         $& $4.7\times 10^{48}$ &$ 9.1\times 10^{27}$ & 0.83\\
4      &$    1.6         $& $3.7\times 10^{49}$ &$ 7.1\times 10^{28}$ & 1.18\\
5      &$1\times 10^6    $& $4.2\times 10^{46}$ &$ 4.6\times 10^{19}$ & 1.50\\
5      &$1.2\times 10^3  $& $2.7\times 10^{48}$ &$ 2.4\times 10^{24}$ & 1.51\\
6      &$2.8\times 10^4  $& $4.4\times 10^{48}$ &$ 1.7\times 10^{23}$ & 1.85\\
6      &$    1.4         $& $3.9\times 10^{49}$ &$ 1.1\times 10^{29}$ & 1.86\\
7      &$3.8\times 10^2  $& $4.4\times 10^{51}$ &$ 1.3\times 10^{28}$ & 2.21\\
7      &$    1.2         $& $2.4\times 10^{49}$ &$ 1.6\times 10^{29}$ & 2.23\\
8      &$    1.2         $& $8.9\times 10^{48}$ &$ 5.0\times 10^{28}$ & 2.55\\
9      &$    1.3         $& $3.7\times 10^{48}$ &$ 1.3\times 10^{28}$ & 2.62\\
10     &$2.5\times 10^5  $& $2.3\times 10^{48}$ &$ 1.1\times 10^{22}$ & 2.90\\
10     &$    8.7         $& $2.6\times 10^{50}$ &$ 3.8\times 10^{28}$ & 2.91\\
\hline
\end{tabular}
\caption{Properties of the ejected mass. Ejected mass in some
pulses are divided into two part: first part with sufficient mass
and move faster; the later part moves slower and cannot collide with
first part. ($M=1.55M_{\odot}$, $\Gamma_n$=1.85 and
$B^{1/4}$=160 MeV).}
\end{minipage}
\end{table*}

\begin{table*}
 \centering
 \begin{minipage}{140mm}
  \begin{tabular}{|c|c|c|c|c|}
  \hline
Pulse & $\Gamma $ & Energy(ergs) & Mass (g) & Time (s) \\
 \hline
1    & $    21          $& $1.6\times 10^{45}$ &$ 8.7\times 10^{22}$ & 0.15\\
2    & $2.6\times 10^2  $& $1.0\times 10^{45}$ &$ 4.4\times 10^{21}$ & 0.47\\
3    & $3.4\times 10^2  $& $3.2\times 10^{45}$ &$ 1.1\times 10^{22}$ & 0.79\\
4    & $1.3\times 10^2  $& $5.3\times 10^{45}$ &$ 4.5\times 10^{22}$ & 1.10\\
5    & $2  \times 10^4  $& $3.6\times 10^{48}$ &$ 2.0\times 10^{23}$ & 1.41\\
6    & $    2.2         $& $6.9\times 10^{49}$ &$ 6.2\times 10^{28}$ & 1.74\\
7    & $    4.8         $& $2.0\times 10^{50}$ &$ 6.0\times 10^{28}$ & 1.89\\
8    & $2.3\times 10^2  $& $3.1\times 10^{48}$ &$ 1.5\times 10^{25}$ & 1.91\\
9    & $1.6\times 10^6  $& $2.9\times 10^{48}$ &$ 2.0\times 10^{21}$ & 2.08\\
9    & $    5           $& $4.7\times 10^{50}$ &$ 1.3\times 10^{29}$ & 2.09\\
10   & $    1.1         $& $4.2\times 10^{48}$ &$ 3.4\times 10^{28}$ & 2.13\\
11   & $    8           $& $1.2\times 10^{46}$ &$ 1.9\times 10^{24}$ & 2.41\\
12   & $    31          $& $2.5\times 10^{44}$ &$ 9.2\times 10^{21}$ & 2.73\\
\hline
\end{tabular}
\caption{Properties of the ejected mass. ($M=1.7M_{\odot}$,
$\Gamma_n$=1.85 and $B^{1/4}$=160 MeV).}
\end{minipage}
\end{table*}

\begin{table*}
 \centering
 \begin{minipage}{140mm}
  \begin{tabular}{|c|c|c|c|c|}
  \hline
Pulse & $\Gamma $ & Energy(ergs) & Mass (g) & Time (s) \\
 \hline
1    & $6.1\times 10^5  $& $9.1\times 10^{45}$ &$ 1.7\times 10^{19}$ & 0.14\\
2    & $    30          $& $2.5\times 10^{46}$ &$ 9.5\times 10^{23}$ & 0.46\\
3    & $    7.8         $& $1.3\times 10^{46}$ &$ 2.1\times 10^{24}$ & 0.76\\
4    & $7.9\times 10^4  $& $1.6\times 10^{47}$ &$ 2.2\times 10^{21}$ & 1.06\\
5    & $    5.9         $& $4.1\times 10^{47}$ &$ 9.3\times 10^{25}$ & 1.37\\
6    & $    14          $& $3.5\times 10^{48}$ &$ 3.1\times 10^{26}$ & 1.68\\
7    & $2.2\times 10^6  $& $2.5\times 10^{48}$ &$ 1.3\times 10^{21}$ & 2.01\\
7    & $    8.2         $& $4.2\times 10^{50}$ &$ 6.4\times 10^{28}$ & 2.01\\
8    & $3.1\times 10^6  $& $4.2\times 10^{48}$ &$ 1.5\times 10^{21}$ & 2.22\\
8    & $    3.9         $& $1.9\times 10^{50}$ &$ 7.4\times 10^{28}$ & 2.22\\
9    & $1.5\times 10^2  $& $1.0\times 10^{45}$ &$ 7.3\times 10^{21}$ & 2.31\\
10   & $    59          $& $7.3\times 10^{43}$ &$ 1.4\times 10^{21}$ & 2.64\\
\hline
\end{tabular}
\caption{Properties of the ejected mass.  ($M=1.8M_{\odot}$,
$\Gamma_n$=1.85 and $B^{1/4}$=160 MeV).}
\end{minipage}
\end{table*}

\begin{table*}
 \centering
 \begin{minipage}{140mm}
  \begin{tabular}{|c|c|c|c|c|}
  \hline
Pulse & $\Gamma $ & Energy(ergs) & Mass (g) & Time (s) \\
 \hline
1    & $2.6\times 10^2  $& $6.1\times 10^{46}$ &$ 2.6\times 10^{23}$ & 0.13\\
2    & $    6.4         $& $2.5\times 10^{47}$ &$ 5.2\times 10^{25}$ & 0.43\\
3    & $6.3\times 10^4  $& $1.4\times 10^{46}$ &$ 2.4\times 10^{20}$ & 0.73\\
4    & $1  \times 10^2  $& $2.4\times 10^{47}$ &$ 2.7\times 10^{24}$ & 1.03\\
5    & $    1.42        $& $1.9\times 10^{49}$ &$ 4.9\times 10^{28}$ & 1.32\\
6    & $4.3\times 10^2  $& $4.9\times 10^{48}$ &$ 1.3\times 10^{25}$ & 1.62\\
6    & $    2.2         $& $7.0\times 10^{49}$ &$ 6.5\times 10^{28}$ & 1.63\\
7    & $1  \times 10^5  $& $3.6\times 10^{49}$ &$ 4.0\times 10^{23}$ & 1.74\\
8    & $9.6\times 10^4  $& $5.2\times 10^{48}$ &$ 6.0\times 10^{22}$ & 1.94\\
8    & $    76          $& $3.7\times 10^{51}$ &$ 5.6\times 10^{28}$ & 1.95\\
9    & $8.8\times 10^3  $& $5.7\times 10^{48}$ &$ 7.2\times 10^{23}$ & 2.25\\
9    & $    2.5         $& $1.9\times 10^{50}$ &$ 1.4\times 10^{29}$ & 2.25\\
10   & $    7.5         $& $2.7\times 10^{47}$ &$ 4.6\times 10^{25}$ & 2.55\\
11   & $    1.2         $& $3.9\times 10^{48}$ &$ 2.0\times 10^{28}$ & 2.86\\
\hline
\end{tabular}
\caption{Properties of the ejected mass.  ($M=1.9M_{\odot}$,
$\Gamma_n$=1.85 and $B^{1/4}$=160 MeV).}
\end{minipage}
\end{table*}

\section{Possible Connection with Gamma-ray Bursts}

\subsection{Duration of mass ejection}

We have pointed out that our numerical simulations are quite limited
by the unphysical numerical damping. However, in realistic
situations, the oscillations triggered by the collapse must also be
damped by some physical dissipation mechanisms. In this section, we
would like to estimate how long the oscillations could last.

First, we note that hydrodynamics effects (e.g., shock waves,
mass-shedding etc.) which can be modeled by our simulations
certainly would play a role in the damping \cite{Lin06, Ab08}. In
particular, the study of \cite{Ab08} suggests that the damping
timescale due to hydrodynamics effects seen in their simulations is
typically a few tens to hundreds ms. Another damping effect which
exists in the case of rotational collapse is that due to
gravitational radiation back-reaction. However, the damping
timescale of this process is much longer than that of the
hydrodynamics effects. The gravitational radiation damping is
negligible and is in fact not taken into account in the works of
\cite{Lin06,Ab08}. There are still other damping mechanisms.
It was first pointed out in \cite{Wang1984} that the dissipation due to
nonleptonic reaction is of great importance and the stellar
pulsations of the quark stars would be strongly damped via $s + u
\leftrightarrow u + d$ in a few milliseconds. On the other hand,
it was shown in \citet{Sawyer1989} and \cite{Madsen1992} that in the
high-temperature limit, which is exactly our case, the bulk
viscosity is dramatically reduced. If we take $m_s \sim 140$ MeV,
$\langle \rho \rangle \sim 10^{15}~{\rm g/cm}^3$ and $\langle T
\rangle \sim 50$ MeV, the damping time scale is $\sim 10$ s
\cite{Sawyer1989}.

If all of the above physical mechanisms cannot efficiently damp out
the oscillations, then the pulse neutrino emission and the produced
mass ejection considered in this paper would be the only processes
to damp out the oscillations. In Tables 1-4 we can see that the
energy carried away by the ejecta in first 3 ms is of the order of
$\sim 10^{50}$ ergs. The escaped neutrinos will carry away
comparable amount of energy. The total oscillation energy is of the
order of $\Delta E_G \sim GM^2 \Delta R/R^2$, where $\Delta R$ is
the change of radius before and after the phase transition. For the
models presented in this paper, $\Delta R/R$ is of the order of
10\%, which gives oscillation energy of the order of several
10$^{52}$ ergs. In other words the oscillations cannot last longer
than several hundreds milliseconds even the neutrino emission and
mass ejection are the only mechanisms to damp out the oscillations.

\subsection{Short GRBs}

There are two kinds of GRBs, i.e., long GRBs with duration larger
than $\sim 2$ s and short GRBs with duration less than $\sim 2$ s
\cite{Kou93, Zh07}. The isotropic $\gamma$-ray energy
released by a short GRB is usually in the range of $10^{49}$ ---
$10^{51}$ ergs, i.e., about two or three orders of magnitude less
than that of long GRBs \cite{Na07}. Assuming that no more than
$\sim$ 10\% of the kinetic energy can be converted to $\gamma$-ray
radiation, then an amount of kinetic energy up to  $10^{50}$ ---
$10^{52}$ ergs should be produced by the central engine. An
example that has a relatively large energy release is GRB 051221A
\cite{Sod06}. The isotropic $\gamma$-ray energy is $1.5
\times 10^{51}$ ergs, and an isotropic kinetic energy of $8.4
\times 10^{51}$ ergs has been estimated for this event \cite{Sod06}.

It is widely believed that long GRBs may
originate from the collapse of massive stars \citep{Wo93}, while
short GRBs may be connected with the merger of binary compact stars
\cite{Eich89,Na07}. However, it is still possible that some GRBs
may be produced by other mechanisms. For example, GRB 060614, a
special nearby long GRB that is not associated with any supernovae,
may be produced by an intermediate mass black hole that captured and
tidally disrupted a star \cite{Lu08}. Another interesting kind of
central engine mechanism involves the phase transition of normal
neutron stars to strange stars \cite{ChDa96, Bo00, Wa00, Ou02}.
Since the phase transition may be processed in a detonative mode,
the details in this process are still largely uncertain and need
further investigation. Especially, numerical simulations are
necessary to help to understand the process.

\subsubsection{Basic equations}

The observations of X-ray, optical and radio afterglows from some
well-localized GRBs have proved their cosmological origin \cite{Co97,Fr97,Ga98,Ge05,Bl06}. The so-called fireball model \cite{Go86, Pa86,  Me92, Re92,Re94,Ka94, Sa96,Sa98}
can basically explain the observational
facts well, and thus it is strongly favored, and widely accepted today. In this
model, the central engine gives birth to some energetic ejecta
intermittently, like a geyser, producing a series of
ultra-relativistic shells. The shells collide with each other at a
radius of $R_{\rm in}$ and produce strong internal shocks. The
highly variable $\gamma$-ray emission in the main burst phase of
GRBs should be produced by these internal shocks. After the main
burst phase, the shells merge into one main shell and continue to
expand outward. It sweeps up circum-burst medium, being decelerated
and producing external shocks. The observed long-lasting and
steadily decaying afterglows (with a much smoother light curve as
compared with the $\gamma$-ray light curve) should be due to these
external shocks.

After examining the numerical results of our simulations, we believe
that the gravitational collapse of a neutron star induced by the
phase transition from normal nuclear matter to quark matter can be
an ideal mechanism for producing GRBs. In this Section, we give some
detailed explanations.

Let us first consider the general case of the collision between two
typical shells. Following the description of \cite{Pa94}, we assume
that the first shell is ejected with a mass of $M_1$, a bulk
velocity of $\beta_1$ and a bulk Lorentz factor of $\gamma_1$. The
second shell is assumed to be ejected after a time interval of
$\Delta t$, with the mass, velocity and Lorentz factor being $M_2,
\beta_2$, and $\gamma_2$, respectively. Here, $\gamma_1 = (1 -
\beta_1^2)^{-1/2}$, $\gamma_2 = (1 - \beta_2^2)^{-1/2}$, and they
satisfy $\gamma_2 > \gamma_1 \gg 1$. The distance between the two
shells is then initially $ c \Delta t$. Since the second shell moves
faster, it will finally catch up with the first shell at a radius of
\cite{Pa94},
\begin{equation}
R_{\rm in} = \frac{\beta_1 \beta_2}{\beta_2 - \beta_1} c \Delta t
     \approx \frac{c \Delta t}{\beta_2 - \beta_1}
     = \frac{2 \gamma_1^2 \gamma_2^2}{\gamma_2^2 - \gamma_1^2} c \Delta t .
\label{rin}
\end{equation}
This is the place where the internal shock appears and the GRB takes
place. At this point, the total elapsed time {\bf \underline{measured in
the static burster frame}} is
\begin{equation}
t_{\rm in} = \frac{R_{\rm in}}{\beta_1 c}
     \approx \frac{\Delta t}{\beta_2 - \beta_1}
     = \frac{2 \gamma_1^2 \gamma_2^2}{\gamma_2^2 - \gamma_1^2} \Delta t .
\label{tin}
\end{equation}
However, due to relativistic effect, the observed elapsed time since
the beginning of the phase transition is only $t_{\rm in} / (2 \gamma_1^2)  \approx
\gamma_2^2 \Delta t / (\gamma_2^2 - \gamma_1^2)$.

After the collision, the two shells will merge into one single shell
and move at a new bulk velocity of $\beta_{\rm bul}$
(correspondingly, with the Lorentz factor $\gamma_{\rm bul}$).
During the process, a portion of the initial bulk kinetic energy
will be dissipated as random internal energy. We denote the average
Lorentz factor of the random internal energy as $\gamma_{\rm i}$.
Since momentum and energy are conserved in the collision, we have \cite{Pa94},
\begin{equation}
M_1 \gamma_1 \beta_1 + M_2 \gamma_2 \beta_2 =
           (M_1 + M_2) \gamma_{\rm i} \gamma_{\rm bul} \beta_{\rm bul},
\label{momentum}
\end{equation}
\begin{equation}
M_1 \gamma_1 + M_2 \gamma_2 = (M_1 + M_2) \gamma_{\rm i} \gamma_{\rm
bul}. \label{energy}
\end{equation}
It is then easy to get the solution for $\beta_{\rm bul},
\gamma_{\rm bul}$ and $\gamma_{\rm i}$ as
\begin{equation}
\beta_{\rm bul} = \frac{M_1 \gamma_1 \beta_1 + M_2 \gamma_2 \beta_2}
              {M_1 \gamma_1 + M_2 \gamma_2},
\label{betabul}
\end{equation}
\begin{eqnarray}
\gamma_{\rm bul} &=& \frac{M_1 \gamma_1 + M_2 \gamma_2}
               {\sqrt{M_1^2 + M_2^2 + 2 M_1 M_2 \gamma_1 \gamma_2 (1 - \beta_1 \beta_2)}}\approx \nonumber\\
               &&\sqrt{\frac{M_1 \gamma_1 + M_2 \gamma_2}{M_1 / \gamma_1 + M_2 /
      \gamma_2}},
\label{gammabul}
\end{eqnarray}
\begin{equation}
\gamma_{\rm i} = \frac{M_1 \gamma_1 + M_2 \gamma_2}{(M_1 + M_2)
\gamma_{\rm bul}}
     \approx \frac{\sqrt{M_1 \gamma_1 + M_2 \gamma_2}
     \cdot \sqrt{M_1 / \gamma_1 + M_2 / \gamma_2}}
     {M_1 + M_2} .
\label{gammai}
\end{equation}
The efficiency of transferring bulk kinetic energy into internal
energy is
\begin{equation}
\epsilon = (\gamma_{\rm i} - 1) / \gamma_{\rm i} = 1 - \gamma_{\rm
i}^{-1}. \label{epsilon}
\end{equation}

When two shells collide, the emission from the shock-accelerated
electrons will correspond to a single pulse in the light curve of
the GRB. The rising time of the pulse is mainly determined by the
time needed for the shocks to cross the two shells. Since the merged
shell is moving toward us ultra-relativistically with a Lorentz
factor of $\gamma_{\rm bul}$, only a small part of the shell (with
an opening angle of $\sim 1/\gamma_{\rm bul}$) will be seen by us.
The decay time of the pulse is then mainly determined by the arrival
time lag of a photon emitted at the angle of $\sim 1 / \gamma_{\rm
bul}$ as compared with the photon emitted simultaneously at the line
of sight. Usually, the decay time is longer than the rising time, so
we can use the decay time to characterize the width of the pulse,
i.e. \cite{Ko97,Pi99}
\begin{equation}
\tau_{\rm pulse} \approx R_{\rm in} / (2 \gamma_{\rm bul}^2 c).
\label{taupulse}
\end{equation}

\subsubsection{GRBs resulting from phase transition?}

According to our simulations, at least more than 10 shells can be
ejected during the phase transition of a neutron star in the first
3 ms. Among these shells, a few will have isotropic energies
larger than $10^{48}$ ergs and expand ultra-relativistically with
Lorentz factors larger than $> 100$ (cf. Tables 1-4).
Note that although our simulations only last for about 3 ms, the
actual duration of the oscillation process may last for several
hundreds milliseconds, as argued at the end of our Section 6.1 .
During this period, tens or even hundreds of ultra-relativistic
shells might be ejected, each with an energy larger than $\sim
10^{48}$ ergs. The total energy enclosed in these relativistic
shells may be $10^{50}$ --- $10^{51}$ ergs.  We suggest that the
collision between these shells can give birth to a GRB.

To describe this process in a quantitative way, we first study the
collision between two specific (but typical) shells. We assume
$M_1 = M_2, \gamma_1= 300, \gamma_2 = 2 \gamma_1 = 600$. We
further assume that the second shell is ejected after a time
interval of $\Delta t = 1$ ms. It is then straightforward to find
that they will collide at a radius of $R_{\rm in} \sim 8
\gamma_1^2 c \Delta t /3 \approx 7.2 \times 10^{12}$ cm, and at
the time of $t_{\rm in} \sim 8 \gamma_1^2 \Delta t / 3 \approx
240$ s. Note that due to relativistic effect, the
observed elapsed time since the beginning of the phase transition
is only $t_{\rm in} / (2 \gamma_1^2)  \sim 4 \Delta t / 3 \approx
1.3$ ms. The merged shell will move at a Lorentz factor of
$\gamma_{\rm bul} \sim \sqrt{2} \gamma_1 \approx 420$, with the
co-moving internal energy characterized by $\gamma_{\rm i} \approx
3/2\sqrt{2} \approx 1.06$. The efficiency of transferring the bulk
kinetic energy into internal energy is $\epsilon = 1 - 1 / 1.06
\approx 5.6\%$. The pulse will have a width of $\tau_{\rm pulse}
\approx 2 \Delta t /3 \approx 0.6$ ms.

In realistic cases, the shells are ejected with variable masses,
variable Lorentz factors, and variable time intervals, respectively.
The conditions then become very complicated. For example, if
$\gamma_2$ is much larger than $\gamma_1$ ($\gamma_2 \gg \gamma_1$,
but  we still assume $M_2 = M_1$), then the solution can be
expressed as $R_{\rm in} \sim 2 \gamma_1^2 c \Delta t$, $t_{\rm in}
\sim 2 \gamma_1^2 \Delta t$, $\gamma_{\rm bul} = \sqrt{\gamma_1
\gamma_2}$, $\gamma_{\rm i} = \sqrt{\gamma_2 / 4 \gamma_1}$,
$\tau_{\rm pulse} = \Delta t \gamma_1 / \gamma_2$. In this case, the
pulse width will be very small. On the other hand, if $\gamma_2$ is
only slightly larger than $\gamma_1$ (still with $M_2 = M_1$), then
$\gamma_{\rm bul} \sim \gamma_1, \gamma_i \sim 1, R_{\rm in}$ and
$t_{\rm in}$ will be very large so that a pulse much wider than
$\Delta t$ can be generated. However, in this case, the energy
transfer efficiency will be extremely low ($\epsilon \ll 1$)
for this single pulse.

In the standard fireball model, it is generally believed that the
total duration of a GRB reflects the active period of the central
engine. In our simulations, the shells are ejected mainly in a few
{\bf hundred }
milliseconds. So, this mechanism is most proper for explaining short
GRBs. The duration may reasonably range between a few ms and several
{\bf hundred }
 ms if the neutrino emission and the  mass ejection are  the
only processes to damp out the stellar oscillations.
It is also possible for the duration to extent to more than 1 s.
For example, if we take $\gamma_1= 350, \gamma_2 = 400$ and
$\Delta t = 300$ ms, then the observed elapsed time for this pulse
will be larger than $t_{\rm in}/ (2 \gamma_1^2) = \gamma_2^2 /
(\gamma_2^2 - \gamma_1^2) \Delta t \approx 4.3 \Delta t \sim 1.3 $
s.  Note that the total isotropic energy of the shells can be
$10^{50}$
--- $10^{51}$ ergs, and this is also consistent with the energy
requirement of  most  short GRBs. If some extent of
anisotropy exists in the process (which is quite possible if the effects of
strong magnetic field of the compact star are further included, see
the discussion in the last paragraph of this section), then the energy
release can even meet the requirement of those rigorous energetic events, such as
GRB 051221A \cite{Sod06}. Also, as noted above, if two
shells are ejected with similar Lorentz factors, they may collide at
a very late time. We further notice from Tables 1 --- 4 that there are many high
energy shells that moves at trans-relativistic speeds (with Lorentz factors
significantly less than $\sim 10$). Late collisions can also be produced
by these energetic shells when they finally catch up with the decelerating
external shock (that produces GRB afterglow). Such late collisions may
manifest themselves as flaring activities (emerged in the afterglow phase),
as frequently observed 1000 --- 10000 s after the trigger of
 GRBs \citep{Bu05,Ro06}.

In our simulations, the shells are ejected periodically, with a
period of $\sim 0.3$ ms. However, it is quite unlikely that we could
observe any obvious periodicity in the $\gamma$-ray light curve. The
reason is that the shells have different masses, velocities, and
energies. The periodicity will then be most likely smeared out at
the time of collision.

The energy releases in our simulations are basically isotropic. The
reason is that we did not include the effects of rotation and
macroscopical magnetic field in our calculations. When these effects
are considered, the energy releases should show some features of
anisotropy. This is an interesting point that needs to be
investigated in further details.

\section{Discussions}

In this paper we have studied the possible consequences of the
phase-induced collapse of neutron stars to hybrid stars. We have
found that both the density and the temperature inside the star will
oscillate with the same period, but almost 180$^{\circ}$ out of
phase, which will result in the emission of intense pulsating
neutrinos. The temperatures at the peaks of pulsating neutrino fluxes
are 10-20 MeV, which are 2-3 times higher than non-oscillating case. Since the
electron/positron pair creation rate sensitively depends on the temperature at
neutrinosphere, the efficiency of pair creation increases dramatically.
 We want to point out that the intense pulse neutrino and pair luminosity can be maintained due to the oscillatory fluid motion, which
can carry thermal energy directly from the stellar core to the surface. This process can replenish the energy loss of neutrino
emission much quicker than the neutrino diffusion process.
Part of neutrino energy,
roughly (1-1/$e$), and pairs inside the star will be absorbed by the matter very near the
stellar surface. When this amount of energy exceeds the
gravitational binding energy of matter, some mass near the stellar surface
will be ejected, and this mass will be further accelerated by
absorbing pairs created from the neutrino and antineutrino
annihilation processes outside the star.
Unlike inside the star, the surface properties, e.g. position of the neutrinosphere,
surface temperature etc., are very sensitive to the surface perturbation. The neutrino and pair
luminosities can be varied from pulse to pulse. This results in the large variation of
the Lorentz factor of the mass ejecta. The internal collisions among these mass ejecta
may produce the short time variabilities of the Gamma-ray Bursts, which can be as short as submilliseconds.
Although mass ejecta are
ejected periodically, each ejecta can have different masses and
Lorentz factors as explained before. Therefore, the intrinsic period could not be
observed. Although we can only simulate the oscillating stars for a few millisecond,
we can speculate that this may be a possible mechanism for short Gamma-ray Bursts based on the following
reasons. \cite{Chan09} have estimated that the viscous damping time for such oscillating system is of order of 10 s. By assuming that if neutrinos and pair emissions are the only damping mechanisms,
the pulsations can last less than $\sim 3s$ \cite{Chan09}, which is roughly the characteristic time scale of short Gamma-ray Bursts.

The phase-transition from a neutron
star to a strange star was simulated in \cite{FrWo98}, with the conclusion that this process is most
likely not a gamma-ray burst mechanism. They mimic the
phase-transition by the arbitrary motion of a piston deep within the
star, and they have found that the mechanic wave will eject $\sim
10^{-2}M_{\odot}$ baryons, which causes the baryon contamination for
the gamma-ray bursts. In our simulations, we assume a sudden change
of equation of state to mimic the phase-transition, and we use the
Newtonian hydrodynamic code to study the response of the stellar
interior, after such a sudden change of the EOS. In our simulations
we find that the mass ejection by the motion of the fluid is very
small. We estimate that the major mass ejection would result from
the heating of neutrinos and pairs on the stellar crust, which is
not modeled in the simulations. Our total energy output and total
mass in ejecta are close to that of \cite{FrWo98} (cf. Tables 1-4).
However, the neutrino energy injection is pulsating, and hence the
mass ejection is also pulsating. The masses of individual ejecta range
from $\sim 10^{-9}M_{\odot}$ to $\sim 10^{-4}M_{\odot}$, with output energy
in the range of $10^{48}$ ergs to $10^{50}$ ergs. Therefore, some ejecta
cannot be relativistic, and they cannot contribute to GRBs.
However,
there are still many relativistic ejecta in each simulation model, which
can have Lorentz factors $>$100, and with a total energy of $\sim
10^{50}$ --- $10^{51}$ ergs.
We conclude that this could be a possible mechanism for short GRBs.

Finally, we want to remark that
our numerical simulations describe a
spherically symmetric, non-rotating,  and collapsing stellar object.
Also the effect of the magnetic fields was not taken into account. Therefore, the
radiation emission produced in this model is isotropic. However, a
realistic neutron star should have finite angular momentum and
strong magnetic field, and hence these two factors could produce
asymmetric mass ejection. This effect will be considered in future
work.

\section*{Acknowledgements}

We thank M.C. Chu, Z.G. Dai, P. Haensel, T. Lu, K.B. Luk, V.V. Usov
and K.W. Wu for useful discussion, and the anonymous referee for very useful suggestions. KSC and TH are supported by the GRF Grants of the Government of the Hong Kong SAR under HKU7013/06P and HKU7025/07P
respectively.  YHF is supported by the National Science Foundation of
China (grant 10625313), and the National Basic
Research Program of China (grant 2009CB824800).  LML is supported by
the Hong Kong Research Grants Council (Grant No. CUHK4018/07P) and
the direct grant (Project IDs: 2060330) from the Chinese University
of Hong Kong. The computations were performed on the Computational
Grid of the Chinese University of Hong Kong and the High Performance
Computing Cluster of the University of Hong Kong.

\end{document}